\documentclass[preprint,12pt]{aastex61}

\pdfoutput=1

\shorttitle{``X-Shaped'' Radio Sources II.}
\shortauthors{Saripalli \& Roberts}
\usepackage{graphicx}

\begin{document}

\title{WHAT ARE ``X-SHAPED'' RADIO SOURCES TELLING US?\\ II. PROPERTIES OF A SAMPLE OF 87}

\author{Lakshmi Saripalli}
\affil{Raman Research Institute, C. V. Raman Avenue, Sadashivanagar, Bangalore 560080, India}

\author{David H.\ Roberts}
\affil{Department of Physics MS-057, Brandeis University, Waltham, MA 02453-0911, USA}

\email{lsaripal@rri.res.in, roberts@brandeis.edu}

\begin{abstract}
In an earlier paper we presented Jansky Very Large Array multi-frequency, multi-array continuum imaging of a unique sample of low axial-ratio radio galaxies. In this paper, the second in the series, we examine the images to know the phenomenology of how the off-axis emission relates to the main radio source. Inversion symmetric offset emission appears to be bimodal and to originate from one of two strategic locations: outer ends of radio lobes (Outer-deviation) or from inner ends (Inner-deviation). The latter sources are almost always associated with edge-brightened sources. With S- and Z-shaped sources being a subset of Outer-deviation sources this class lends itself naturally to explanations involving black hole axis precession. Our data  allow us to present a plausible model for the more enigmatic Inner-deviation sources with impressive wings; as for outer-deviation sources these too require black hole axis shifts although also requiring plasma backflows into relic channels. Evolution in morphology over time relates the variety in structures in inner-deviation sources including XRGs. With features such as non-collinearities, central inner-S ``spine'', corresponding lobe emission peaks, double and protruding hotspots not uncommon, black hole axis precession, drifts, or flips could be active in significant fraction of radio sources with prominent off-axis emission. At least 4\% of radio galaxies appear to undergo black hole axis rotation. Quasars offer a key signature to recognize rotating axes. With rich haul of sources that have likely undergone axis rotation  our work shows the usefulness of low axial-ratio sources to pursue searches for binary supermassive black holes.

\end{abstract}

\keywords{black holes --- galaxies: active --- radio continuum: galaxies}

\section{INTRODUCTION}
\label{s:Introduction}

Notwithstanding dramatic bends in jets seen in wide- and narrow-angle tail sources or the abrupt bends in quasar jets or possible jet flailings at lobe ends (inferences in support of the dentist's drill model for lobe advances; \cite{S82}), jets have long been renowned for their directional stability and integrity. Connecting cores to hotspots several hundred to even megaparsecs away, timescales of up to $10^8$ yr have been inferred as the length of time (ages of Giant Radio Galaxies (GRGs), \cite{MJS}) for which jet axes can remain unperturbed. 

While these phenomenal jet characteristics put the spotlight on jet generation and jet confinement mechanisms, observations of radio sources also revealed a different aspect to jets -- that besides directional stability and bending, jets can precess. Quite early it was realized that the structures of some radio galaxies seemed to require the jet axis to have rotated clockwise or anti-clockwise to understand inversion or rotationally symmetric emission on large scales seen on either side of the radio core or host galaxy (e.g., NGC\,315, 3C\,315, \cite{Betal79, M80}). Lack of constancy in jet axis direction should not surprise us considering that the very innards of active galaxies, the accretion disk-black hole systems and their environs, are not particularly quiescent regions given the presence of gas inflows, rotating disks, disk flow instabilities, twisting magnetic fields, disk winds, precessing and warping disks, black hole realignments and counter-alignments, among other processes. 

That the jet or black hole axis can remain unperturbed for a long time in such dynamic conditions is in itself indeed remarkable. Puzzlingly however, the more frequently expected disturbances suffered by the black hole axis, while receiving fair attention in theoretical exercises (e.g., \cite{NP,KLOP}), have not been the subject of systematic observational campaigns. Such studies over the years have been largely of serendipitous nature involving instances of individual sources that appeared to require rotation of jet axis to understand the radio structures \citep{E1978, Betal79, Betal81, G1982, C1984, Hu1984, A1985, W1987, DT2002,  DVVOM, S2008a, Biju, RDKH, Na17}. 

With radio jets believed to form along AGN black hole spin axes, the defining axes of radio galaxies (whether traced by the line formed by hotspots at lobe ends or by the jets themselves) serve as proxy for the black hole axis. Happily, the relatively large electron radiative lifetimes help in keeping visible various morphological features in the synchrotron plasma deposited by the jets and this helps tracing the temporal history of the black hole spin axis. Radio galaxies thus form a valuable tool in leading one to perturbed black holes in AGN. While black hole axis changes in radio galaxies result in deposition of radio synchrotron plasma in regions offset from the jet axis, the offset features hold within them clues to details of the axis perturbation itself. Whether the jets continued their plasma transport as they rotated, speed of rotation, angle described, power of the jets, angle to the line of sight all can affect the morphology of offset emission regions generated. Examining the full range of axis perturbations and their characteristics are useful ingredients in study of supermassive black hole spin evolution and hence, in black hole mass growth rate through cosmic times.

On the other hand though, uncertainties abound as regards the very linking of offset radio emission with BH axis shifts where the offset emission can have an altogether different physical origin, possibilities that are famously captured by the so-called X-shaped radio galaxies (XRGs). Indeed a number of physical mechanisms have been suggested for the formation of XRGs. While on one hand an extreme perturbation to the black hole axis, a ``spin flip'' is invoked in one of the scenarios to explain the X-shaped structures seen in a small fraction of radio galaxies \citep{M2002, DT2002}, separately, the role of host thermal halos in giving rise to off-axis emission has been cited by several authors \citep{Le1984, Wo95, C2002, SS2009, HC2010}. Other suggested mechanisms include jet-optical shell interactions and twin AGNs with axes oriented at near orthogonal directions \citep{GBGW,LRP}. Moreover, for the less spectacular but commonly seen central inversion symmetric extensions in the general radio galaxy populations, while the origin is not clear, interactions with host thermal halos have been implicated \citep{Le1984,SS2009}. Supporting the role of macro flows of synchrotron plasmas in generating the central wings are radio-optical studies that are beginning to hint at correlations between radio axis orientations and the projected axes of the host ellipticals as a function of the type, power and size of manifested AGN processes \citep{C2002,SS2009,BB2009,GCR16}. 

Theoretical understanding requires quantifying such correlations as tests for proposed models to identify likely physical processes operating in these radio galaxies. It is important to identify such sources that may be used as test-beds for investigating mechanisms such as deflection of back flowing synchrotron plasma in thermal halos on one hand and jet axis rotation on the other hand. In \cite{Paper0}, we presented our preliminary efforts at examining the physical processes responsible for off-axis radio emission based on categories of low-axial ratio source morphologies where we characterized the variety in morphological distortions based on our analysis of archival VLA data for a subsample of the sources from the original 100-source sample of \cite{C2007}. In \cite{Paper1} we presented results from a comprehensive set of EVLA observations that we proposed for the entire sample of 99 low-axial ratio sources (one source of clear head-tail morphology from \cite{Paper0} was excluded). 

A complete sample such as we have imaged would obviously encompass within it a variety of causative processes for location of off-axis radio emission. In this paper with the much improved images for previously observed sources and new images for the remaining, we substantially extend our previous research to a better categorization of low-axial ratio source morphologies.  This is a first step towards identifying and quantifying via sample studies the variety and characteristics of axis change phenomena and their occurrence rates. The polarization and spectral index information on the source sample will be separately presented in future works.


The plan of this paper is as follows: The selection of the sample of 87 that we have chosen for the study is discussed in Section~\ref{s:Sample}. The properties of our working sample are discussed in Section~\ref{s:Results} and Section~\ref{s:Discussion}. Sections~\ref{s:O-Dev} and \ref{s:I-Dev} discuss the outer and inner deviation sources. Source with two apparent independent set of axes are discussed in Section~\ref{s:Independent}. In Section~\ref{s:quasars} we discuss the special role of quasars in this study. Section~\ref{s:AxisChange} discusses the  variety of parameters to consider in relating observed morphologies to axis rotation. In Section~\ref{s:Model} we present a discussion of our results and a tentative model to understand offset emission that is not obviously due to black hole axis precession. The paper is summarized in Section~\ref{s:Summary}. Section~\ref{s:Future} briefly discusses potential future directions for this research.

The paper is based entirely on analysis of morphologies of a unique class of radio galaxies, the low axial ratio sources. Given the central role of radio morphologies in the work presented here we advise that the reader have access to our first paper, \cite{Paper1}, while following the presentation in this paper.

\section{THE SAMPLE}
\label{s:Sample}

\cite{Paper1} presents every image made of the 99 sources from the \cite{C2007} sample, both  archival images from \cite{Paper0} and our Jansky VLA observations. Not all sources could be observed in all desired arrays and bands due to the priority ``C'' scheduling. Altogether there are 236 total intensity images. The sources were observed over a year from February 2016 through January 2017 at 
L- and S-band in the A-, B-, and C-arrays (one epoch was in the transitional AB-array; see \cite{Paper1} for more details). The resolution we obtained for 
the images in L-band is $4\arcsec$ (B-array) whereas in S-band the resolutions range from $2\arcsec$ to $6\arcsec$ (AB-, B- and C- arrays).

It is important to point out that our interest in the sample is to identify and understand processes shaping radio galaxies without the dominant effect of the environmental influences. We therefore do not consider sources where environment is known to have a dominant influence such as in wide angle tail sources and narrow angle tail sources. We examined all available multi-band images of the 99 sources to prune the original sample of such ``contaminant'' sources. In addition we discarded sources that upon examination of the new data revealed them to be either groups of independent radio sources or simply a lone lobe of a larger radio galaxy. In all we excluded 13 such sources. The 87 remaining sources formed our working sample (Table~1 and 2). The sources excluded from the sample are listed in Table~3, along with reasons for their rejection. 

Tables~1 and 2 present details for each of the sources in our sample including those not observed in the present observational program. As our observations have provided higher resolution than that available with the FIRST maps \citep{C2007} we have examined and revised, where possible, the host galaxy identifications based on our new data. In the few cases where we had no high-resolution imaging available we relied on images in \cite{C2007} for source classification.

The columns from 1 to 6 are self explanatory. Columns give 7 -- 9 give information on morphological characteristics based on examination of 
all available images for each of the 87 sources. Sources for which we can trace the origin of the off axis emission to the inner ends of lobes we
label ``I-dev'' (that stands for Inner-deviation type sources), those where the origin is traced to outer ends of lobes are labeled ``O-dev'' (for Outer-deviation type sources) 
and those for which the classification was unclear are labeled ``I-dev?'' and ``O-dev?.'' Finally sources where the off-axis emission could not be traced to either inner or 
outer ends of the lobes and instead formed a central swath of emission across the main source axis are classified under ``XRG?'' column (see below for more details). 
Column 11 gives host galaxy apparent r magnitude and notes on emission lines, and column 12 gives the projected linear size based on the radio images and the redshifts. 
 Source total powers at 1.4~GHz are given in column~13. Where available we used our L-band B-array data and when a source is not observed in this band and array
we use as first priority our S-band C-array data and as second priority our S-band B-array data (when both are available) using a spectral index of $\alpha = 1$ 
($S_\nu \propto \nu^{-\alpha}$) to extrapolate to 1.4~GHz. When neither is available we used the FIRST fluxes to obtain the powers.
Columns 14--16: here we give properties related to whether the off-axis emission in I-dev sources have sharp inner edges and whether they are separated by emission gaps 
(column 14), asymmetry of the main source given as ratio of longer lobe extent to shorter lobe extent (column 15) and information on whether the source has hotspots 
and whether the stronger hotspot is associated with the longer wing lobe (column 16). 


\startlongtable
\begin{deluxetable*}{llllllccc}
\tablecaption{Sample Sources 1/2 \label{tab:sample1}}
\tablecolumns{9}
\tablehead{ \colhead{IAU} & \colhead{RA} & \colhead{DEC} & \colhead{$z$\tablenotemark{a,b}}  & \colhead{FR, LAS\tablenotemark{c}} & \colhead{Optical ID?} & \colhead{XRG?\tablenotemark{d}} & \colhead{I-dev?} & \colhead{O-dev?}  \\ \colhead{Name} & \colhead{J2000} & \colhead{J2000} & \colhead{}  & \colhead{} & \colhead{Radio Core?} & \colhead{} & \colhead{} & \colhead{Axis Change?} \\ \colhead{(1)} & \colhead{(2)} & \colhead{(3)} & \colhead{(4)} & \colhead{(5)} & \colhead{(6)}  & \colhead{(7)} & \colhead{(8)} & \colhead{(9)}  }
 \startdata
J0001$-$0033 &00 01 40.18 & $-$00 33 50.6&0.247&FR-II, 35\arcsec&Y, Y&&&Y, Y \\
J0036+0048&00 36 36.21 & +00 48 53.4&0.591&FR-II, 41\arcsec&Q, Y&&Y& \\
J0045+0021&00 45 42.11 &+00 21 05.5&0.433*&FR-II, 32\arcsec&Y?, Y&&Y&\\
J0113+0106 &01 13 41.11 &+01 06 08.5&0.281&FR-II, 177\arcsec&Y,Y&&Y& \\
J0115$-$0000&01 15 27.37 & $-$00 00 01.5&0.381&FR-II, 96\arcsec&Y,Y&&Y&\\
J0144$-$0830&01 44 09.98 & $-$08 30 02.8&0.181&FR-I, 39\arcsec&Y, N&Y?&& \\
J0145$-$0159&01 45 19.99 & $-$01 59 47.9&0.1264&FR-I, 56\arcsec&Y, Y&&&Y, -- \\
J0147$-$0851&01 47 19.28 & $-$08 51 19.6&0.455&FR-II, 100\arcsec&Q, Y&&Y& \\
J0211$-$0920&02 11 46.96 & $-$09 20 36.6&0.236*&FR-II, 60\arcsec&Y, Y&&Y?&Y?, -- \\
J0225$-$0738 &02 25 08.62 & $-$07 38 49.1&0.659&FR-II, 66\arcsec&Y, Y&&Y& \\
J0702+5002&07 02 47.92 &+50 02 05.3 &0.0946&FR-I, 34\arcsec& Y, Y&&Y& \\
J0725+5835 &07 25 32.27 &+58 35 27.4 &\dots&FR-II, 80\arcsec&fnt, Y&&Y& \\
J0805+4854&08 05 44.0 &+48 54 58 &\dots&FR-II, 30\arcsec&fnt, Y&&Y& \\
J0813+4347 &08 13 00.11 &+43 47 48.5 &0.128&FR-I, 42\arcsec&Y, Y&&&Y, -- \\
J0836+3125&08 36 35.46&+31 25 51.2&0.376&FR-II, 65\arcsec&Y, Y&&Y& \\
J0838+3253 &08 38 44.61 &+32 53 11.8 &0.213&FR-II, 59\arcsec&Y, Y&&&Y, -- \\
J0845+4031 &08 45 08.40 &+40 31 15.4 &0.429&FR-II, 37\arcsec&Y, Y&&&Y, -- \\
J0846+3956&08 46 03.58 &+39 56 57.9 &0.257*&FR-II, 35\arcsec&Y, Y&&Y& \\
J0859$-$0433 &08 59 50.19 & $-$04 33 06.9&0.356&FR-II, 53\arcsec&Y, Y&&Y& \\
J0914+1715&09 14 05.21 &+17 15 54.3 &0.52&FR-II, 61\arcsec&Q, Y&&Y&N, Y? \\
J0917+0523 &09 17 44.31 &+05 23 10.1 &0.592&FR-II, 57\arcsec&Y, Y&&Y?&\\
J0924+4233&09 24 46.98 &+42 33 47.1 &0.227&FR-II, 62\arcsec&Y, ?&&Y?&Y?, -- \\
J0941$-$0143&09 41 22.58 & $-$01 43 01.1&0.383&FR-II, 39\arcsec&Y, Y&&Y& \\
J0941+2147 &09 41 58.31 &+21 47 44.2 &0.572*&FR-II, 31\arcsec&Y, N&&Y& \\
J0943+2834 &09 43 02.25 &+28 34 45.8 &0.574&FR-II, 96\arcsec&Y, Y&&Y& \\
J1005+1154 &10 05 52.53 &+11 5437&0.166&FR-II, 43\arcsec&Y, Y?&&Y?&Y? \\
J1008+0030&10 08 11.43 &+00 30 00&0.0977&FR-I/II, 80\arcsec&Y, Y&&&Y?, -- \\
J1015+5944 &10 15 41.14 &+59 44 45.2 &0.527&FR-II, 48\arcsec&Q, Y&&Y&\\
J1040+5056&10 40 22.48 &+50 56 25.1 &0.154&FR-II, 63\arcsec&Y, Y&&Y?&Y? \\
J1043+3131&10 43 18.63 &+31 31 06.1 &0.036&FR-II, 32\arcsec&BLRG, Y&Y?&&\\
J1049+4422 &10 49 35.28 &+44 22 04.0 &0.477*&FR-II, 60\arcsec&Y, Y&&Y& \\
J1054+5521&10 54 00.60 &+55 21 53 &\dots&FR-II, 51\arcsec&N, N&&Y& \\
J1055$-$0707&10 55 52.56 & $-$07 07 19.1 &\dots&FR-II, 84\arcsec&Y, N&&Y& \\
J1102+0250&11 02 06.56 &+02 50 45.7 &\dots&FR-II, 51\arcsec&Y, Y&&&Y, Y \\
J1114+2632&11 14 27.10 &+26 32 58.8 &0.66*&FR-II, 27\arcsec&Y, Y&&&Y, -- \\
J1120+4354 &11 20 16.18 &+43 54 51.4 &0.612&FR-II, 48\arcsec&Y, Y&Y?&&N, Y? \\
J1135$-$0737&11 35 36.78 & $-$07 37 00.1 &0.602&FR-II, 39\arcsec&Y, Y&&&Y, Y \\
J1140+1057&11 40 49.70 &+10 57 56.3 &0.081&FR-II, 73\arcsec&Y, Y&&Y& \\
J1200+6105&12 00 06.00&+61 05 44.5 &0.294*&FR-II, 60\arcsec&Y, Y&&Y?& \\
J1202+4915&12 02 35.10 &+49 15 31.7 &0.688*&FR-II, 40\arcsec&Y, Y&&Y& \\
J1206+3812&12 06 17.35 &+38 12 34.9 &0.838&FR-II, 43\arcsec&Q, Y&&Y& \\
J1207+3352 &12 07 32.92 &+33 52 40.1 &0.079&FR-II, 57\arcsec&Y, Y&&&Y, -- \\
J1210$-$0341 &12 10 18.77 & $-$03 41 53.3 &0.178&FR-II, 49\arcsec&Y, Y&Y?&Y?& \\
J1211+4539&12 11 02.50 &+45 39 14.4 &0.698*&FR-II, 44\arcsec&Y, N&&Y& \\
J1218+1955&12 18 59.15 &+19 55 28.0 &0.424&FR-II, 76\arcsec&Y, Y&&Y& \\
J1227$-$0742&12 27 12.5 & $-$07 42 00&\dots&FR-II, 70\arcsec&N, Y?&&Y?&0, Y? \\
J1228+2642&12 28 03.83 &+26 42 26.7 &0.201&FR-II, 48\arcsec&Y, Y?&----&----&0, Y? \\
J1232$-$0717&12 32 52.50 & $-$07 17 28.9 &\dots&FR-II, 45\arcsec&Y, Y&&&0, Y? \\
J1247+4646 &12 47 26.84 &+46 46 06.0 &0.838*&FR-II, 36\arcsec&Y, Y?&Y?&& \\
J1253+3435 &12 53 24.29 &+34 35 18.2 &0.358&FR-II, 44\arcsec&Y, Y&&&Y, -- \\
J1258+3227 &12 58 32.87 &+32 27 40.8 &\dots&FR-II, 61\arcsec&Y, Y&&Y& \\
J1309$-$0012&13 09 49.74 & $-$00 12 35.6 &0.419&FR-II, 70\arcsec&Y, Y&&Y& \\
J1310+5458 &13 10 15.40 &+54 58 34.2 &0.355&FR-II, 31\arcsec&Q Y&&&Y, Y? \\
J1316+2427 &13 16 38.27 &+24 27 32.4 &0.447&FR-II, 39\arcsec&Y, Y&&Y& \\
J1327$-$0203 &13 27 38.22 & $-$02 03 09.9&0.183&FR-I, 45\arcsec&Y, Y?&Y?&&Y?, -- \\
J1342+2547 &13 42 45.32 &+25 47 11.5 &0.585&FR-II, 35\arcsec&Q, Y&&&Y, -- \\
J1345+5233&13 45 41.64 &+52 33 35.6 &0.4*&FR-II, 29\arcsec&Y, Y&Y?&Y?& \\
J1348+4411&13 48 04.59 &+44 11 24.2 &0.266&FR-II, 37\arcsec&Y, Y&&&Y, -- \\
J1351+5559 &13 51 42.14 &+55 59 43.1&0.069&FR-I, 68\arcsec&Y, Y&&&Y, - \\
J1353+0724 &13 53 06.37 &+07 24 37.1 &0.316*&FR-II, 77\arcsec&Y, N&&Y& \\
J1406$-$0154&14 06 48.61 & $-$01 54 16.5&0.64&FR-II, 61\arcsec&Y, Y&&Y& \\
J1406+0657&14 06 02.41 &+06 57 15.7 &0.549&FR-II, 24\arcsec&Q, Y?&Y&& \\
J1408+0225&14 08 28.41 &+02 25 50.3 &0.179&FR-II, 25\arcsec&BLZ, Y&&&0, Y? \\
J1411+0907&14 11 21.14 &+09 07 35.4 &0.194*&FR-II, 89\arcsec&Y?, N&&Y?& \\
J1424+2637&14 24 40.53 &+26 37 30.5 &0.037&FR-II, 145\arcsec&Y, Y&&&Y, -- \\
J1430+5217 &14 30 17.34 &+52 17 35.0 &0.368&FR-II, 38\arcsec&Q, Y&&&Y, -- \\
J1433+0037&14 33 51.87 &+00 37 23.4 &0.504&FR-II, 61\arcsec&Y, Y&&Y?&Y?, -- \\
J1434+5906&14 34 02.17 &+59 06 53.3 &0.538&FR-II, 33\arcsec&Y, Y&Y?&& \\
J1437+0834 &14 37 38.90 &+08 34 22.1 &0.316*&FR-II, 79\arcsec&Y, Y&&Y& \\
J1444+4147&14 44 07.25 &+41 47 50.3 &0.188&FR-II, 93\arcsec&Y, Y?&&Y& \\
J1454+2732 &14 54 42.22 &+27 32 11.8 &0.692*&FR-II, 55\arcsec&Y, N&&&Y? \\
J1455+3237&14 55 58.28 &+32 37 32.5 &0.084&FR-II, 49\arcsec&Y, Y&&Y?& \\
J1456+2542&14 56 43.16 &+25 42 17.5 &0.536&FR-II, 42\arcsec&Y, N&&Y?& \\
J1459+2903&14 59 41.95 &+29 03 31.8 &0.146&FR-II, 86\arcsec&Y, Y&&Y& \\
J1501+0752&15 01 57.44 &+07 52 26.7 &0.659&FR-II, 74\arcsec&Y, Y&&Y?&Y? \\
J1515$-$0532 &15 15 33.66 & $-$05 32 51.0&\dots&FR-II, 86\arcsec&Y, Y&&Y& \\
J1537+2648&15 37 07.33 &+26 48 23.2 &0.287&FR-II, 70\arcsec&Y, Y&&&Y, -- \\
J1600+2058&16 00 38.95 &+20 58 51.7 &0.174&FR-II, 73\arcsec&Y, Y&&Y& \\
J1603+5242 &16 03 44.96 &+52 42 20.6 &0.292&FR-II, 95\arcsec&Q, Y&&Y&\\
J1606+0000 &16 06 12.68 &+00 00 27.1 &0.059&FR-II?, 35\arcsec&Y, Y&&&N, Y? \\
J1606+4517&16 06 38.88 &+45 17 37.1 &0.556&FR-II, 48\arcsec&Y, N&&Y& \\
J1614+2817&16 14 28.40 &+28 17 31.0 &0.107&FR-II, 43\arcsec&Y?, Y &Y?&& \\
J1625+2705&16 25 30.68 &+27 05 46.3 &0.525&FR-II, 32\arcsec&Q, Y&&&Y?, -- \\
J1653+3115&16 53 37.31 &+31 15 27.5&0.652&FR-I, 58\arcsec&Q, Y&&&Y, -- \\
J1655+4551&16 55 16.2 &+45 51 08 &0.478*&FR-II, 30\arcsec&Y, Y?&&Y?& \\
J1656+3952&16 56 36.65 &+39 52 58.3 &0.3&FR-II, 32\arcsec&Y, Y&&&Y, -- \\
J2359$-$1041&23 59 46.47 & $-$10 41 15.4&0.216*&FR-II, 73\arcsec&Y, Y&&&Y?, -- \\
\enddata
\tablenotetext{a}{Redshifts are largely from SDSS DR-12, with the remaining ones from Cheung (2007), Cheung et al. (2009), and Landt et al. (2010).} 
\tablenotetext{b}{A * is for photometric redshifts from SDSS DR12.}
\tablenotetext{c}{ \ Largest linear size along main source.} 
\tablenotetext{d}{Is the source a candidate ``true XRG?''}
\end{deluxetable*}

\startlongtable
\begin{deluxetable*}{lllclcl}
\tablecaption{Sample Sources 2/2 \label{tab:sample2}}
\tablecolumns{7}
\tablehead{ \colhead{IAU} & \colhead{$m_r$, Em. Lines?} & \colhead{LLS (kpc)}  & \colhead{ $P_{1.4}$\tablenotemark{a} } & \colhead{Gaps/Edges?\tablenotemark{b}} & \colhead{Asymmetry\tablenotemark{c}} & \colhead{Hs? Shs-LW-side?\tablenotemark{d,e}}  \\ \colhead{(10)} & \colhead{(11)} & \colhead{(12)} & \colhead{(13)} & \colhead{(14)} & \colhead{(15)}& \colhead{(16)} }
 \startdata
J0001$-$0033 	&	17.4, N	&	137	&	25.05	&	N2/N2	&	1.4	&	N, 0 \\
J0036+0048	&	20.3, Y,strong	&	276	&	26.59	&	Y1/Y1	&	1.3	&	2dsim, N \\
J0045+0021	&	23.4, NA	&	182*	&	26.52	&	Y1/Y1	&	1	&	2dsim, 0 \\
J0113+0106 	&	18.5, Y,strong	&	GRG, 759	&	25.62	&	Y1/N2	&	1.3	&	2dsim, Y \\
J0115$-$0000	&	20.4, Y, NA	&	503	&	25.97	&	N1/Y1	&	1.1	&	2dsim, Y \\
J0144$-$0830	&	18.6, N	&	120	&	24.67	&		&	1.1	&	0, 0 \\
J0145$-$0159	&	15.9, NA	&	128	&	24.97	&		&	1.3	&	0, 0 \\
J0147$-$0851	&	20.2, Y,strong	&	585	&	26.30	&	Y1/N	&	1.5	&	2dsim, Y \\
J0211$-$0920	&	18.56, NA	&	227*	&	25.38	&	Y2/N	&	1.4	&	2dsim, Y \\
J0225$-$0738 	&	20.9, Y,strong	&	465	&	26.76	&	Y1/Y1	&	1.5	&	2dsim, N \\
J0702+5002	&	15.5, Y,strong	&	60	&	24.84	&	Y1/Y1	&	1.2	&	Nhs, 0 \\
J0725+5835 	&	19.9, NA	&	?	&	\dots	&	N/N	&	1.2	&	2dsim, Y \\
J0805+4854	&	22.5, NA	&	?	&	\dots	&	N/N	&	1.4	&	2dsim, Y \\
J0813+4347 	&	15.98, Y,weak	&	97	&	25.10	&		&	1.8	&	Nhs, 0 \\
J0836+3125	&	19.43, Y,weak	&	339	&	26.00	&	N2/N2	&	1.1	&	2dsim, Y \\
J0838+3253 	&	16.94, Y, weak	&	206	&	25.05	&		&	1.3	&	2dsim, 0 \\
J0845+4031 	&	19.04, Y,strong	&	211	&	26.01	&		&	1.1	&	2dsim, N \\
J0846+3956	&	20.7, NA	&	141*	&	25.64	&	N2/N2	&	1.2	&	2sim, 0 \\
J0859$-$0433 	&	NA, Y,strong	&	267	&	25.83	&	N2/N2	&	1.4	&	2dism, 0 \\
J0914+1715	&	19.71, Y,strong	&	383	&	27.16	&	Y1/Y1	&	1.7	&	2dsim, Y \\
J0917+0523 	&	20.34, Y,strong	&	380	&	26.88	&	N2/Y1	&	1.2	&	2nsim, Y \\
J0924+4233	&	17.75,Y,weak	&	226	&	25.63	&	?/N2	&	1.1	&	N, 0 \\
J0941$-$0143	&	19.03, Y,ok	&	204	&	26.59	&	Y2/Y1	&	1.1	&	2nsim, 0 \\
J0941+2147 	&	23.1, NA	&	205*	&	26.45	&	Y1?/Y1?	&	1.3	&	2dsim, Y \\
J0943+2834 	&	20.6, Y,strong	&	634	&	26.87	&	Y1/Y1?	&	1.3	&	2dsim, 0 \\
J1005+1154 	&	16.3, N	&	124	&	25.06	&	Y1/Y1	&	2	&	2sim, 0 \\
J1008+0030	&	15.83, NA	&	146	&	24.92	&		&	1.1	&	2dsim, Y \\
J1015+5944 	&	18.24, Y,strong	&	302	&	26.46	&	N2/-	&	1.6	&	2dsim, N \\
J1040+5056	&	17.1, Y,weak	&	170	&	25.28	&	Y2/N2	&	1	&	2sim, 0 \\
J1043+3131	&	16.23, Y,strong	&	23	&	24.49	&		&	1.6	&	2dsim, 0 \\
J1049+4422 	&	20.13, NA	&	361*	&	26.15	&	Y2/Y1	&	1.5	&	2dsim, No \\
J1054+5521	&	?, NA	&	?	&	\dots	&	0/0	&		&	2dsim, 0 \\
J1055$-$0707	&	17.6, NA	&	?	&	\dots	&	poor resln	&	1	&	2dsim, 0 \\
J1102+0250	&	22.3, NA	&	?	&	\dots	&		&	1.1	&	2dsim, Y \\
J1114+2632	&	20.95, NA	&	192*	&	26.77	&		&	1.9	&	2dsim, 0 \\
J1120+4354 	&	20.39, Y,strong	&	328	&	26.87	&	N, N	&	1	&	2dsim, 0 \\
J1135$-$0737	&	19.2, N	&	264	&	26.18	&		&	1	&	2dsim, 0 \\
J1140+1057	&	15.68, Y,ok	&	112	&	24.80	&	N2/Y1	&	1	&	2dsim, Y \\
J1200+6105	&	18.32, NA	&	266*	&	25.84	&	N2/N2	&	1.2	&	2dsim, 0 \\
J1202+4915	&	22.45, NA	&	290*	&	26.34	&	Y1/Y1	&	1.1	&	2dsim, Y \\
J1206+3812	&	18.35, Y,strong	&	332	&	26.85	&	N2/N2	&	1.2	&	2nsim, 0 \\
J1207+3352 	&	15.13, Y,strong	&	85	&	24.91	&	Y1/N2	&	1.1	&	2dsim, 0 \\
J1210$-$0341 	&	17.8, N	&	148	&	25.20	&	N2/N2	&	1.2	&	2dsim, 0 \\
J1211+4539	&	22.16, NA	&	316*	&	26.71	&	0/0	&	0	&	2dsim, N \\
J1218+1955	&	19.89, Y,strong	&	425	&	26.77	&	N2/N2	&	1.2	&	2dsim, 0 \\
J1227$-$0742	&	NA, NA	&	?	&	\dots	&	N2/N2	&	1	&	0, 0 \\
J1228+2642	&	17.82, N	&	160	&	25.32	&	0, 0	&	1.8	&	0, 0 \\
J1232$-$0717	&	19.7, NA	&	?	&	\dots	&	0, 0	&	2.1	&	2dsim, 0 \\
J1247+4646 	&	22.2, NA	&	277*	&	26.64	&	0, 0	&	1	&	2nsim, 0 \\
J1253+3435 	&	19.15, Y,strong	&	223	&	26.08	&	Y1/N	&	1.1	&	2dsim, 0 \\
J1258+3227 	&	17.0, NA	&	?	&	\dots	&	Y1/Y1	&	1.2	&	2dsim, Y \\
J1309$-$0012	&	19.6, Y,strong	&	391	&	27.00	&	Y1/Y1	&	1.1	&	2sim, 0 \\
J1310+5458 	&	19.2, Y,strong	&	154	&	26.04	&	0/0	&	1.1	&	2dsim, Y \\
J1316+2427 	&	19.58, Y,strong	&	226	&	25.76	&	Y1/N	&	1	&	2dsim, Y \\
J1327$-$0203 	&	16.62, Y,strong	&	140	&	26.15	&	0/0	&	1.1	&	0, 0 \\
J1342+2547 	&	20.15, Y,ok	&	233	&	26.52	&	N2/N2	&	2	&	2dsim, 0 \\
J1345+5233	&	20.7, NA	&	157*	&	25.67	&	0/0	&	1.2	&	2nsim, 0 \\
J1348+4411	&	18.35, Y,strong	&	154	&	25.68	&	0/0	&	1.2	&	2dsim, N \\
J1351+5559 	&	15.03, v. weak	&	90	&	24.49	&	0/0	&	1.2	&	0, 0 \\
J1353+0724 	&	19.3, NA	&	356*	&	25.99	&	N2/0	&	1.1	&	2dsim, N \\
J1406$-$0154	&	20.9, Y,strong	&	425	&	27.04	&	N2/Y1	&	1.6	&	2dsim, Y \\
J1406+0657	&	18.53, Y,strong	&	155	&	26.62	&	0/0	&	1.2	&	2dsim, 0 \\
J1408+0225	&	18.45, Y,strong	&	75	&	25.90	&	0/0	&		&	0, 0 \\
J1411+0907	&	20.58, NA	&	275*	&	25.45	&	Y1/N2	&	1.5	&	2nsim, 0 \\
J1424+2637	&	13.8, Y,strong	&	107	&	24.20	&	Y1/N2	&	1.1	&	0, 0 \\
J1430+5217 	&	19.42, Y,strong	&	196	&	26.58	&	Y2/Y2	&	?	&	2sim, 0 \\
J1433+0037	&	20.03, Y,strong	&	378	&	26.43	&	N2/Y1	&	1	&	2dsim, 0 \\
J1434+5906	&	20.1,prspctrm	&	211	&	26.62	&	N/N	&	1.1	&	2dsim, 0 \\
J1437+0834 	&	19.18, NA	&	367*	&	25.84	&	0/0	&	1.1	&	2nsim, N \\
J1444+4147	&	17.3, Y,v. weak.	&	293	&	25.67	&	0/0	&	1	&	2sim, 0 \\
J1454+2732 	&	21.5, NA	&	359*	&	26.66	&	0/0	&	1.4	&	2dsim, Y \\
J1455+3237	&	15.9, N	&	78	&	24.04	&	N2/N2	&	1.4	&	0, 0 \\
J1456+2542	&	20.6, N	&	269	&	25.62	&	N2/N2	&	1.5	&	0, 0 \\
J1459+2903	&	16.46, N	&	222	&	25.32	&	0/0	&	1.1	&	2sim, 0 \\
J1501+0752	&	20.9, Y,strong	&	522	&	26.98	&	N2/N2	&	1.2	&	2dsim, Y \\
J1515$-$0532 	&	19.9, NA	&	?	&	\dots	&	N2/N2	&	1.2	&	2dsim, N \\
J1537+2648	&	18.29, Y,strong	&	305	&	25.92	&	0/0	&	1	&	2dsim, 0 \\
J1600+2058	&	17.17, Y,strong	&	217	&	25.68	&	0/0	&	1	&	0/0 \\
J1603+5242 	&	17.9, NA	&	417	&	26.04	&	N2/N2	&	1.1	&	2dsim, Y \\
J1606+0000 	&	14.95, NA	&	40	&	25.30	&	0/0	&	1	&	2dsim, 0 \\
J1606+4517	&	21.7, Y,strong	&	320	&	26.20	&	0/0	&	1.1	&	0, 0 \\
J1614+2817	&	15.94, N	&	84	&	25.18	&	0/0	&	1.1	&	0, 0 \\
J1625+2705	&	18.5, Y,strong	&	202	&	26.87	&	0/0	&	1.5	&	2dsim, 0 \\
J1653+3115	&	20.8, Y,strong	&	407	&	26.68	&	0/0	&	1	&	0, 0 \\
J1655+4551	&	20.5?, NA	&	177*	&	25.88	&	0/0	&	1.5	&	2dsim, Y \\
J1656+3952	&	18.82, Y,strong	&	141	&	25.75	&		&	1	&	2nsim, 0 \\
J2359$-$1041	&	18.9, NA	&	258*	&	25.51	&	0/0	&	1.4	&	2dsim, 0 \\
\enddata
\tablenotetext{a}{Log$_{10}$(power at 1.4~GHz, W~Hz$^{-1}$).}
\tablenotetext{b}{Gaps between inner ends of lobes/whether inner ends have sharp edges: 1 \& 2 indicate only one or two gaps and edges. 0 indicates - not applicable.}
\tablenotetext{c}{Ratio of longer lobe extent to shorter lobe extent along main source.} 
\tablenotetext{d}{Are there hotspots? (2dsim = two dissimilar strength hotspots; 2nsim $=$ two nearly similar strength hotspots; 2sim $=$ two similar strength hotspots.} 
\tablenotetext{e}{Is the stronger hotspot on the side of the longer wing?}
\end{deluxetable*}




\begin{deluxetable*}{clll}
\tablecaption{Rejected Sources \label{tab:reject}}
\tablecolumns{4}
\tablehead{ \colhead{Name} & 
 \colhead{$z$} & \colhead{Morphology} &  \colhead{Notes}  }
\startdata
J0033$-$0149 & 
0.1301 & WAT? &   \tablenotemark{a} \\
J0049+0059  & 
 0.3044 & confused & \tablenotemark{b} \\
J0143$-$0119 & 
0.520,& FR-I,  HT? &   \tablenotemark{c} \\
J0821+2922 & 
 (0.332) & lobe\arcmin? &   \tablenotemark{d} \\
J1111+4050 & 
0.0737& HT &  \tablenotemark{e}  \\
J1128+1919 & 
 \dots & HT &     \tablenotemark{f} \\
J1201$-$0703 & 
?? & WAT &   \tablenotemark{g}  \\
J1210+1121 & 
0.1963  & WAT? &  \tablenotemark{h} \\
J1227+2155 & 
0.459  &  confused  &   \tablenotemark{i}  \\
J1330$-$0206  & 
 0.0866 & WAT &   \tablenotemark{j} \\
J1339$-$0016 & 
0.1452 & WAT &   \tablenotemark{k} \\\
J1522+4527 & 
 (0.450) & ?  &   \tablenotemark{l}\\
J2226+0125 & 
   (0.216) & galaxy group &   \tablenotemark{m} \\
\enddata
\tablenotetext{a}{Complex structure. WAT projected? CD host. Dense group of galaxies.}
\tablenotetext{b}{Complex. Superposed sources?} 
\tablenotetext{c}{NS wingsize $=$ 540~kpc} 
\tablenotetext{d}{DR12-photoz. SW lobe of much larger very asymmetric~3.5\arcmin\ FR-II. } 
\tablenotetext{e}{Head-Tail source.}
\tablenotetext{f}{Head-Tail source with core and ID at extreme south.} 
\tablenotetext{g}{Not in DR12. }
\tablenotetext{h}{ \ Likely severely projected.}
\tablenotetext{i}{DR12. Cluster relic?}
\tablenotetext{j}{ \ Likely a WAT?} 
\tablenotetext{k}{ \ Likely a WAT. Many galaxies in vicinity of host.} 
\tablenotetext{l}{DR12. Nature of src unclear. Given ID is wrong--no ID for this src.  Both ``wings'' are on same side. No host though.} 
\tablenotetext{m}{DR12. No spectrum. Two galaxies seen in radio.}
\end{deluxetable*}





\section{RESULTS: SAMPLE CHARACTERISTICS}
\label{s:Results}

\subsection{Overview}

\begin{figure}[htb!]
\plotone{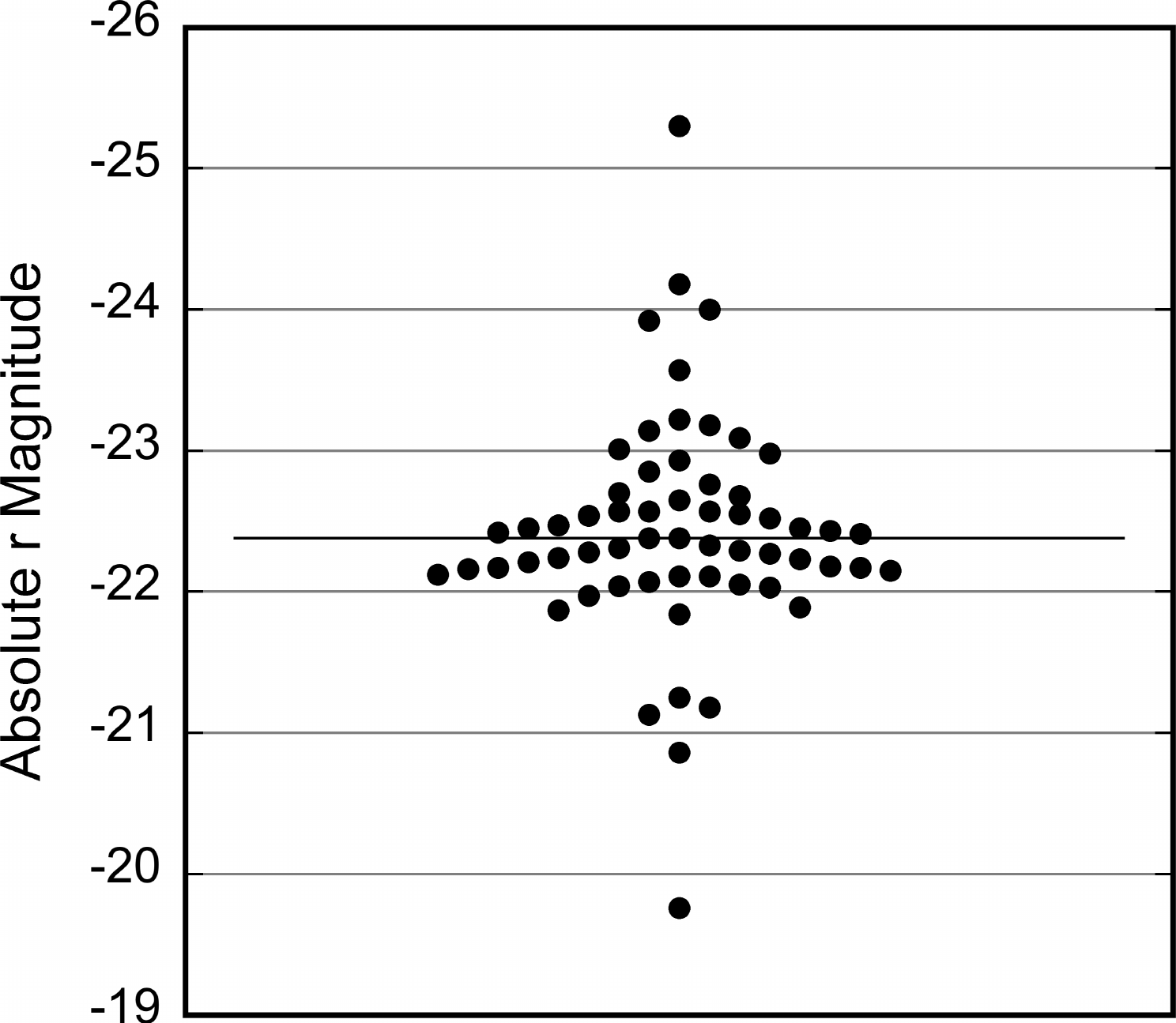}
\caption{Distribution of absolute r magnitudes for those sources in the sample with spectroscopic redshifts. The horizontal line through the points is the mean.\label{fig:Mr}}
\end{figure}

The break up of this 87-source sample is as follows: seven sources are classified as of FR-I type morphology and 78 sources as FR-II type morphology \citep{FR74}. For two sources for which we are unsure of the classification we label them as FR-I/II and FR-II? type. The majority of sources being classified as edge-brightened FR-II type is expected given that the low-axial ratio sample of 100 sources has been selected from the FIRST survey, itself a flux-density limited sample. The sample of 87 sources has 12 confirmed quasars (14\%) with five others that show radio morphological signatures\footnote{Bright cores, one-sided jets, and asymmetries in lobe extents and hot spot strengths.} of being inclined at small angles to the line of sight, although these need spectroscopic confirmation. Besides the 12 quasars one source is identified with a BLRG host and one other is identified as a blazar. Most of the sample sources have optical identifications (85/87) with a large fraction (68/87) having clear detected radio cores. In those without detected radio cores the optical identifications are clear as in J0144$-$0830 although several also have faint optical objects located in the central regions as in J1456$+$2542.   A large fraction of the IDs, 72\% (61/85) have confirmed redshifts (this excludes the 17 with only photometric redshifts; we have used optical information from SDSS-DR12 and \cite{C2007,C2009,L10}). In all nine sources do not have redshift information (whether spectroscopic or photometric). All redshifts are below $z=1$ (the highest redshift is $z=0.838$). 54 sources have optical spectra available of which as many as 70\% (38/54) have strong emission lines. Six sources have weak emission lines and 10 have only absorption line spectra. The distribution of absolute r magnitudes for the sample is shown in Figure~\ref{fig:Mr}. The mean and median $M_r$ are both about $-22.4$. Median projected linear size of the sample is 217~kpc (using only sources with spectroscopic redshifts) and the sample includes a giant radio galaxy of projected size 759~kpc. 

\epsscale{0.5}
\epsscale{1.0}

\subsection{Division By Off-Axis Morphology}

Sources displaying off-axis emission regions that clearly originate from the inner ends of lobes (the I-dev sources) form the largest fraction of the sample (43\%, 37/87 sources). The sources with off-axis emission regions clearly originating from the outer ends of the lobes (the ``O-dev'' sources) constitute only 22\% (19/87). For seven sources we recognize a central off-axis swath of emission that cannot be clearly traced to either lobe, these we label "XRG?". There are several sources where we are unclear as to their classification whether of inner or outer (lobe end) deviation type sources and we indicate them with a question mark (seven of I-dev? type and four of O-dev? type). There are also nine where the structures are confused between I-dev, O-dev as well as XRG type (marked as pairing between I-dev?, O-dev?, and XRG?). In addition there are four sources (J1228+2642, J1232-0717, J1408+0225 and J1606+0000) where we recognize possible axis change (because they exhibit either multiple pairs of opposing emission peaks and/or clear inner S structure)   but without clear classification between cases of offset emission from either inner or outer lobe ends . 

\subsection{F-R Characteristics of the Various Classes}

Remarkably all but one of the 37 sources of I-dev type are of FR-II morphology. Such a clear result is not the case for those of the O-dev type, where four out of nineteen or one-fifth are instead of FR-I type. Given the dominance of FR-II sources in the sample it is expected that this morphological type is also the dominant source type among the I-dev and O-dev sources, although the near absence of FR-I source type among the I-dev sources as compared to the O-dev sources is suggestive of a close connection between FR-II and I-dev type sources. This close connection between I-dev type offset emission and FR-IIs is also seen when we consider that as many as 46\% of FR-IIs (36/78) are of I-dev type where as O-dev type FR-IIs form only 19\% (14/78). On the other hand, interestingly, with 4/7 FR-Is having O-dev type offset emission and only one (1/7) associated with I-dev type (J0702+5002) it appears that O-dev type offset emission is more closely connected with FR-I type morphologies than with FR-II type although this needs a larger sample to support such an association. With I-dev type sources almost exclusively associated with FR-II type sources FR-IIs appear to more easily form inner deviation type sources compared to FR-Is. 

There are seven sources where we have indicated a possible XRG classification. These are sources where there is a transverse band of emission situated about the core or host and which we cannot clearly link to either one of the lobes. Six of these are of FR-II morphologies, although the seventh, J0144-0830 is also likely be a FR-II with 
hotspots at the lobe ends as indicated by the L-band A-array archival map \citep{Paper0}. 

What is the nature of the 20 sources (excluding XRG type) where the classification is unclear? For the nine sources for which classification is given as either ``I-dev, O-dev, or I-dev,  XRG or O-dev, XRG'' type several factors give rise to uncertainty. In sources with low axial ratio lobes the off-axis emission is often not very distinct to allow for easy classification (e.g., J0211-0920, J0924+4233, J1005+1154 and J1040+5056). In some others the likely projection of the off-axis emission onto the main source causes classification to be unclear for one lobe with the other lobe having clear inner-lobe deviation (e.g., J1433+0037, J1501+0752). In a few others the resolution is too coarse for classification (J1210-0341, J1327-0203, J1345+5233). In the remaining doubtful cases (seven I-dev? type and four O-dev? type sources) as well as the four sources where we have not been able to assign any classification, higher resolution and sensitivity are needed for better imaging the structures, which may clarify the structure classification. 

\begin{figure}
\plotone{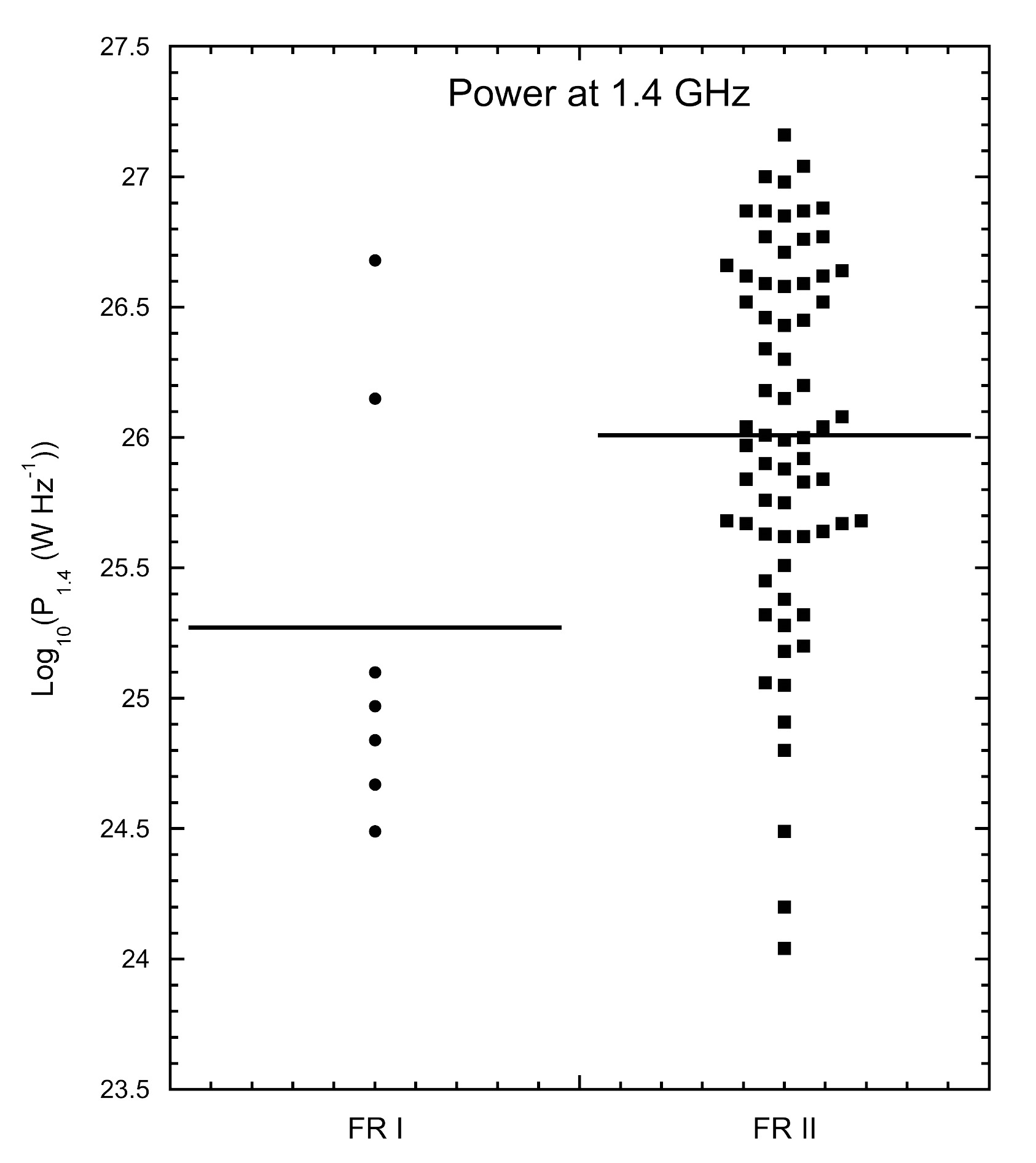}
\caption{The distribution of radio powers of the sources in our sample with  redshifts. They are divided according to their FR morphologies. \label{fig:powers}}
\end{figure}

Using our multi-frequency data we computed radio powers of our sample sources  (described in Section~\ref{s:Sample} above). The distribution of L-band total powers in the sample is given in Figure~\ref{fig:powers},  where the sources are divided according to their FR classifications.

\epsscale{0.45}

\subsection{Sources with Large Wings}

Several sources in the sample (11/37 I-dev sources) have wings that are of same extent as, or longer than, the associated lobes. Two I-dev sources (J0702+5002 and J0859$-$0433) have both wings with extents longer than their respective lobes, six I-dev type sources have one of the two wings with extent longer than the respective main lobe and three I-dev sources have one of the two wings having same extent as the associated main lobe. Four sources classified as XRGs (J1043+3131, J1406+0657, J1434+5906 and J1614+2817) have the central orthogonal swath of emission with extent longer than the main source extent. 

Considering only the I-dev sources we looked at the fractional extent of the wings with respect to their associated lobes for each of the two lobes of a source. We report a tendency for the shorter lobe of a source to be associated with fractionally longer wing. For seven non-symmetric (asymmetry ratio $> 1.2$) I-dev sources (out of the eleven that have at least one of the wings longer than, or of equal extent to, its respective lobe) six have their longer wings associated with the shorter lobes. To widen the exercise we examined the remaining I-dev type sources that have wing extents smaller than their respective lobes; for the 12 among them that are non-symmetric (asymmetry ratio $>1.2$; and excluding in addition two that have no ID or radio core) we find a large fraction of sources (9/12, or two-thirds of the sources) have fractionally longer wings on the side of the shorter lobes. 

We report a tentative trend for the stronger hotspot in I-dev sources to be associated with the lobe with a longer wing. Of the 21 I-dev sources with dissimilar strength hotspots and wings with dissimilar extents, in 14 the stronger hotspot is associated with a lobe that is also associated with the longer wing. 

\subsection{S, Z, and Restarted Sources}
\label{s:s+z}

Among the 87 sources there are 12 sources with clearly S or Z-type structures (nine have FR-II classification and three have FR-I classification; Table~\ref{tab:sample1}). In at least 21 sources (that include three S- and Z-type sources) we recognize an inner-S shaped central emission ``spine'' linking the hotspots and the core. Non-collinearity of hot spots or lobe extremities with the core is not uncommon in these sources with five of S and Z-type sources and eight of the 20 sources with inner-S spine exhibiting non-collinearity. We recognize several sources as candidates for axis change (see section~\ref{s:Independent}) because of corresponding emission peaks in each lobe (besides outer hotspots in the lobes) or a distinct radio source sharing the same radio core but along a different axis (J0001$-$0033, similar to 3C\,293; \cite{Betal81}. We also recognize one source, J1227$-$0742 where a combination of symmetrically located emission regions and protruding hotspots suggest an axis change such as examined in the case of the GRG B0707-359; \cite{Sa2013}. Non-collinearities, inner S-shaped spine, protruding hotspots, double hotspots, corresponding inner-emission peaks (besides outer hotspots in the lobes) are features present in the sample sources in addition to the recognized off-axis emission often linked to the inner or outer ends of lobes. 

The sample has six sources in which the AGN may have restarted its activity (J0813+4347, J0924+4233, J1040+5056, J1327-0203, J1424+2637, J1459+2903). We recognize and classify a source to be ``restarting''  if the lobes are relaxed with only warm-spots with a relatively bright core with a jet (e.g., J1327$-$0203) or straddled by close twin emission peaks.


\section{RESULTS: DISCUSSION}
\label{s:Discussion}
 
Radio morphological clues are already well acknowledged in having given a lead in recognizing that radio AGN cycle through active, dead and active-again phases (e.g., \cite{Cetal92,SSH,Setal2000}, and the many others following these initial works). To this AGN behavior is now added, via radio morphological analysis, the additional possibility of jet axis that can undergo perturbations even as the nuclear activity is ongoing and also, between independent activity phases, as has been long recognized via individual source studies (see Introduction section for a list of these works). 

The select radio source sample we have used and the multi-frequency radio images that we have obtained in total and polarized intensity reveal a vast range of source morphologies covering both radio galaxies and quasars as well as both FR-types. Such a comprehensive sample study would cover a range of axis perturbations and here we attempt to fathom the perturbation type that may have led to the variety of radio structures we have observed. The thrust of our study is to ultimately examine physical causes underlying deposition of synchrotron plasma into regions significantly offset from the axis of a radio source and to this end we present here an analysis of source morphologies where we have attempted to classify the different types of off-axis structures, and to examine them in the context of the main previously proposed models. With this as objective we do not engage below in a comparison between various reigning models for X-shaped radio galaxies and for now suggest previous works that have ably made comparative and qualitative analysis (e.g., \cite{SS2009, GBGW, HKR2012}). In the discussion below we also do not inquire into causes of jet axis drifts or precession or flips for which we await more quantitative analysis in our future works.

At the outset, one outstanding property we recognize for our sample of low-axial ratio sources is the mostly inversion symmetric nature of the structures Ð whether in inner-deviation type sources or outer-deviation type. The selection criterion of the original \cite{C2007} sample did not favor this property specifically. The inversion symmetric structures appear to hint at links with a host-located property that dominates the population of low axial ratio sources. 

A sample such as this, selected on the basis of low axial ratios and selected for hints of extended structure characteristic of wings, is vulnerable to effects of projection, de-selecting sources where off-axis emission is in a plane perpendicular to the sky plane. Therefore, the sample will include a higher fraction of sources where the axis of rotation may have been more or less in the sky plane or close to it. As a result, there is also a greater chance of sources with relatively smaller angles of precession being picked up as members of the sample than would otherwise be possible.

\section{THE O-DEV TYPE SOURCE STRUCTURES}
\label{s:O-Dev}

The radio galaxy literature, through serendipitous discoveries, has long recognized S and Z-shaped radio galaxies, which in the manner of classification of our low-axial ratio source sample fall in the outer-deviation or O-dev type source category, where the inversion symmetric off-axis deviations to the source structure originate from the outer ends of lobes. Earlier works on S and Z-shaped radio galaxies inferred rotation of jet axes in accounting for the peculiar inversion symmetric structures. There are 19 O-dev type sources in the sample of which 12 are S- and Z-type source structures. While the precise conditions that help generate the S- and Z-shape may not necessarily all be operating in the remaining, one can however infer an axis rotation as causing the similarly O-dev nature of the off-axis emission in these sources just as inferred in the S- and Z-shape sources. As noted in Section~\ref{s:s+z} features such as non-collinearities, central inner-S 'spine', corresponding inner emission peaks in the two lobes, double hotspots as well as protruding hotspots are not uncommon in the sources in the sample. These are all likely manifestations of a changing jet axis. 

If O-dev sources represent those where the black hole axes have experienced a rotation then given the detection rate of O-dev type sources in our sample we infer at least 1\% (19/1648,  where 1648 is the original number of sources from which the XRG candidates were drawn \citep{C2007}) of radio galaxies appearing to undergo black hole axis rotation. The lower limit considers the fact that the minimum rotation angle above which all rotations will be detected will increase progressively with increasing angle of the plane of rotation with respect to the plane of sky. While large angle rotations at increasingly large angle to the sky plane will be seen as small-angle rotations in the O-dev sources those with smaller rotation angles will be missed. Therefore if we include sources that are discriminated against, those for which axis rotations happen in a plane orthogonal to the sky plane as well as the unseen rotations in rotation planes in between, this fraction will be higher.

Given that the sources are all drawn from a uniformly observed sample the largest rotations we detect will have a higher probability of being in the sky plane. 
That should give us an estimate of the fraction of radio galaxies that undergo large-angle rotations. If we use the nine S and Z-shape sources that cover the largest angle as representative of sources that suffered large-angle rotation in the sky plane then it is only 0.5\% -- amounting to just one out of 200 radio galaxies undergoing a large-angle rotation in the plane of the sky.


\section{THE PECULIAR CASE OF I-DEV TYPE SOURCE STRUCTURES}
\label{s:I-Dev}

While it is easier to account for the morphology of  the obviously ``Z'' or ``S'' type sources or O-dev type sources or ones where we can distinguish two radio sources at an angle to each other, the same is not the case with I-dev type sources. In this latter category there are ones with clear emission gaps in the central region between the two lobes and those where there is no gap. Several of previously recognized X-shaped sources fall into the former category of inner-deviation type sources with emission gaps between lobes. These are also the sources which \cite{GBW2003} and \cite{Z2005} recognized as XRGs having inner "Z-symmetric" wings and who speculated on merger-related oppositely rotating gas flows causing the jets to be bent (also as a result of rotating optical ``shells'' \cite{GBGW}). \cite{Paper0} had separately recognized ``genuine'' XRGs as those sources, which had a distinct central emission feature oriented at a large angle to the main radio axis that could not be traced to be originating from either lobe, essentially to point to the difficulty in applying the same simplistic rapid spin-flip model to the two ``XRG'' types (those with and without gaps). This inner emission-gap feature of some XRGs (I-dev sources in our terminology) was also discussed by \cite{SS2009} as well as \cite{HKR2012} (for more on this see Section~\ref{s:Model} below). The ``Z-symmetric'' wings or ``emission gap'' features seen in some XRGs or the I-dev sources as referred to in this paper therefore have been the focus of some attention in putting together a scenario to explain the peculiar X-structures in a fraction of radio sources. \cite{Go2011} have suggested that inner-deviation type sources too can be understood as a consequence of precessing jets affected by light travel time effects. 

In Tables~\ref{tab:sample1} and \ref{tab:sample2} we note presence and absence of emission gaps as well as presence or absence of distinct edges to inner ends of lobes for all our sample sources. Although, as mentioned above this feature has received attention in the context of XRGs, it has previously been noted in studies of radio galaxies. What docks the radio lobes at the central regions was a question raised decades ago by \cite{JS76} when they observed a fraction of sources exhibiting regions between lobes that lacked emission. With central distortions a common feature in complete radio galaxy samples \citep{Le1984} where the lobes are distorted away from the axis at central regions, emission gaps as well as inner edges to the lobes appear to be commonly ocuring. In their Figure~6 two possibilities were suggested for the formation of lobe inner distortions of inversion-symmetric nature (classified here as I-dev sources): backflows deflected in opposite directions in a ``non-aligned spheroidal gas distribution'' and another where the backflows get channeled into cavities from a past activity along a different position angle. In another model involving backflows, \cite{C2002} argued for escape of synchrotron plasma from a pressured region along directions of steep pressure gradient close to the host minor axis direction, in creating the wings (although this did not address the question of emission gaps). More recently, \cite{HKR2012} examined scenarios involving both of the above models for explaining the central emission gap between the large wings of NGC~326. With regard to the spin-flip model (that opposes the role of backflows in the origin of inversion symmetric central distortions), the very combination of lobe edges and emission gaps was used as an argument against the case where the transverse emission features or wings are viewed as emission remaining from a past activity with axis in that direction \citep{SS2009}.

In the following sections we examine the spin-flip scenario more closely in the context of I-dev sources  with large wings. Here we need to be aware that the lobes created earlier would be subject to (1) local conditions dominant around the host (such as buoyancy in host coronal gas just as is invoked as a possibility for inner lobe regions of radio galaxies and one that we see evidence of in central cluster radio galaxies) (2) synchrotron age-related effects and (3) ``entrainment'' whether of the IGM or entrainment of new synchrotron plasma itself from backflows in the currently active lobes (option ``c'' in Figure~6, \cite{Le1984}). Hence we may not expect that we see symmetric and perfectly located {\em old} lobes of the kind seen in a typical radio galaxy but ones that have seen the ravages of time.

\section{SOURCES EXHIBITING SIGNATURES OF TWO INDEPENDENT RADIO AXES OR PAIR OF EMISSION PEAKS IN EACH LOBE}
\label{s:Independent}

Besides O-dev type structures hinting at change in axis orientation our observations have revealed another sub-sample of sources with other features in them suggestive of axis change. In one of the sources J0001$-$0033, B-array L-band image reveals a second and more extended radio source associated with the EW radio source (and sharing the same radio core), which is oriented at $35^\circ$  position angle and nearly twice in extent. The source has 3C\,293-type separate extended source associated with the host galaxy. Then again J1227$-$0742 has an inner pair of extended ``lobes'' along an independent axis, resembling B0707$-$359 \citep{Sa2013}. A larger sub-group of 14 sources (10 of which have I-dev type off-axis emission and six also show an inner-S shape) has sources with a second pair of emission peaks in each lobe, in some cases forming an axis with the core that is separate from the main radio axis of the source. The additional pair of peaks may be located at the leading ends in which case they are mostly of relatively compact nature ( e.g., J1140+1057, J1253+3435) or they may be located closer in where the peaks are more extended (e.g., J0846+3956).  

Although in some of these sources there is the possibility of fortuitous location of corresponding emission peaks along a separate axis through the core, accompanying features such as inner-S (linking the main hotspots and core) and O-dev type emission trails not only hint at likely axis rotation in these sources but may also be revealing non-smooth nature of the axis change. As in Cygnus-A where not only two activity epochs are identified along two separate axes but the jets in the two epochs are inferred to have also precessed \citep{S2008a,S2008b}, could the jet axes in at least some of these sources have suffered combination of different axis-disturbance types? In a given source there is little reason why the jet axis always describes only one type of perturbation -- that of a steady drift or steady precession or a single flip. Depending on the physical cause a disturbed jet could suffer a combination of perturbations over time and some of them might be more amenable to manifestations as visible features in the extended lobes, e.g. O-dev type off-axis trails with corresponding intensity peaks in them along a separate axis through the core would suggest the drifting axis might have stopped for long enough to leave an imprint before drifting again to then create the source along the main axis. Of course several sources may need better imaging observations to confirm and to better classify the features. Our as yet unexamined spectral index data will be useful in revealing if indeed the features along separate axes in J0001$-$0033 and J1227$-$0742 have spectral index distributions distinct from those of the main sources.

\section{UNIQUE ROLE OF QUASARS IN AXIS ROTATION STUDIES}
\label{s:quasars}

It is worth examining the special role quasars have in deciphering morphologies and in relating them to AGN behavior. As is well known, at orientations close to the line of sight bulk relativistic flows directed towards the observer will get boosted in flux and as a result the core and the approaching jet will appear bright, with the receding jet de-boosted leading to the classic signatures of bright cores and one-sided jets in quasars. If the hotspots also have relativistic advance speeds the approaching hotspot will be brighter than the receding hotspot. Also, in a growing source, light-travel time effects can leave an imprint on the source morphologies in the form of pronounced asymmetry in lobe extents (photons from the farther lobe will always lag behind photons from the closer lobe). Given the unique nature of our sample, composed of low-axial ratio radio sources, we examined our quasar subsample of 12 sources for light-travel time effects with regard to axis rotation. As reasoned above, the more distant lobe will be imaged at an earlier stage relative to the closer or approaching lobe. Just as the age-asymmetry of the two quasar lobes translates into an asymmetry of lobe extents it is expected that it also has an effect on the collinearity of the two lobes if the quasar axis suffers rotation. The more distant lobe axis is expected to be skewed from the axis formed by the approaching lobe and core in a manner such that its hotspot is displaced towards the same side as its associated off-axis emission feature.
 
In the quasar subsample there are five O-dev type sources. We identify approaching lobes in the quasars using a combination of two of the following signatures: presence of brighter and more compact hotspot, presence of a jet, longer lobe extent. Four O-dev type quasars (J1310+5458, J1342+2547, J1430+5217, and J1653+3115) are not only non-collinear they also have the counter-jet side or the weaker, less compact hotspot or the shorter lobe displaced off axis to the same side as the outer deviation trail associated with that lobe. This is just as noted above as what is expected if the quasar axis suffered a rotation. For the fifth O-dev type quasar, J1625+2705, with its inner features surrounded by compact halo-like emission, it requires higher resolution imaging to reveal its structure more clearly. Although it is significantly non-collinear and displays strong projection effects and shows signature of axis rotation of having an inner-S (as well as inner-Z), with the data we have we are not able to fathom its structure and therefore we do not include it in the study.

We take this morphological characteristic seen in four of the five quasars of O-dev type as strong support for axis rotation in these sources, lending confidence in using O-dev type structures to indicate axis rotation. We also examined the five radio sources in our sample whose structures show characteristics similar to that seen in quasars but which lack optical spectra. Three of these are of O-dev type (J1102+0250, J1114+2632 and J2359-1041) of which the same characteristic is seen in all three of the  counter-jet side being skewed towards the same side as the off-axis extension associated with it. In J1114+2632 we used only the lower resolution CS map because the higher resolution BS map is not of sufficient quality for this purpose.  Extending the test to all O-dev sources with jets, there are two sources (1327$-$0203, 1351+5559, neither is a quasar) for which although jet sidedness is clear no effect is seen where as for three others (0838+3253, 1424+2637, 1656+3952, none of which is a quasar) where although jets are seen the sidedness is unclear.

Quasars have historically been associated with sample studies that drew inferences of axis precession \citep{G1982, Hut1988, Betal94} including the more recent MOJAVE VLBI studies 
\citep{Lis13} which indicate at least 30\% AGN jets revealing clues to axis precession, whether of small angle or large angle precession. In \cite{P1993} where the VLA snapshot images of the sample quasars are published, we note non-collinearity of the quasars as the one common feature of all those identified as having S-type structures by \cite{Hut1988}. 

 We also note the interesting aspect of high occurrence rate of O-dev type structures among quasars in our sample. While 42\% of quasars (5/12) are of O-dev type only 19\% (14/73) of the remaining 73 sources (excluding the BLRG J1043+3131 and the blazar J1408+0225) have O-dev type structures.

\section{AXIS CHANGE PARAMETERS TO CONSIDER IN RELATING OBSERVED MORPHOLOGIES TO AXIS ROTATION}  
\label{s:AxisChange}

In relating radio morphologies with axis change phenomenon earlier works have already recognized the influence of different kinds of axis directional changes, projection and light travel time effects. In the next few paragraphs we outline possible ways in which axis changes may be manifest in the extended radio morphologies of radio sources.
 
The axis change type invoked to understand XRGs is the well known, nearly orthogonal, rapid ``spin flip'' of the black hole spin axis \citep{M2002}. The angle over which the directional change happens is large and it is a one-time change. In this kind of axis change, if the AGN remained active all during the flip then the spin flip would have to happen rapidly to account for the lack of continuous emission connecting the two lobe pairs (option ``C'' in Figure~7, \cite{DT2002}).

On the other hand the AGN could have ceased its activity before the axis orientation changed (option ``D'' in Figure~7, \cite{DT2002}). In this case, with the axis change happening between two independent activity episodes, the change can be either rapid (a flip) or slow (a drift) and also the axis change could have happened either soon after the first AGN ceased its activity or a long time after it ceased. Moreover, the new activity episode could have been triggered soon after the axis change or a long time after the axis change. To add further time parameters, the new source may be observed at a young age or at a mature age or after the activity ceased and it became a relic! The axis direction change moreover could be over a large angle or a small angle. 

In addition, the axis change can also be of a precessing nature Ð where the black hole spin axis describes a cone (option ``B'' in Figure~7, \cite{DT2002}), the kind of axis change identified in the galactic micro-quasar, SS\,433. Here too, as explored in previous works on axis precession, the appearance of the resulting radio source can be affected by the angle over which the precession cone is described and its orientation with respect to the line of sight, as well as the speed of precession. 

In all the above types of axis changes the radio source morphology will ultimately also depend on the power of the jets and hence whether FR-I type or FR-II type. It is also possible for the axis to suffer a combination of axis-change types, combining say a drift over a certain angle after which it flips over a small angle among many other possibilities. The parameter space to consider in accounting for the low-axial ratio radio structures is, therefore, large.

\begin{center}
\begin{figure}[h]
\includegraphics[width=4in]{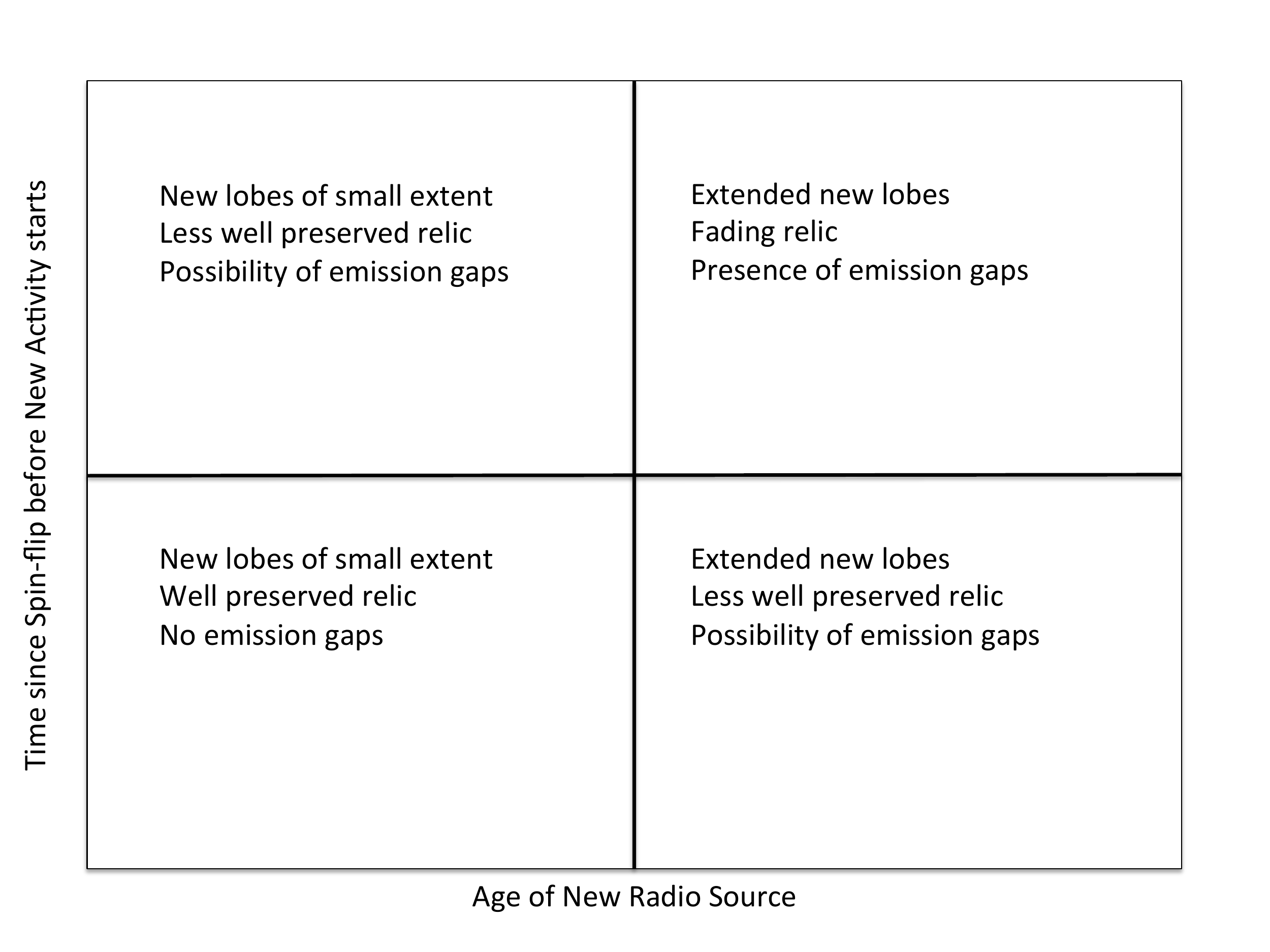}
\caption{How source history affects morphology  in a spin-flip scenario. \label{fig:chart}}
\end{figure}
\end{center}

In Figure~\ref{fig:chart} we present a simplistic chart to indicate morphologies expected in the spin-flip case of an AGN hosting a radio galaxy, as a function of time since spin-flip before the new activity started (y-axis), and age of the new radio galaxy (x-axis). Sources that were created only a short time after axis flip will occupy the lower half of the chart, with new {\em young} sources appearing to the lower left and older sources appearing to the lower right. On the upper half of the chart will be the sources which get created a long time after the spin flip with the upper left corner occupied by the new young sources and the upper right by the older sources. The most divergent and hence more distinguishable morphologies occur for the lower-left and upper-right quadrants.

\section{A GENERAL MODEL TO UNDERSTAND OFFSET EMISSION}
\label{s:Model}

We offer a possible scenario for the most dominant source type in our sample -- the inner-deviation type sources -- weighing in results from this work and building on arguments for and against popular models for XRGs discussed in previous works, particularly in \cite{SS2009}. This discussion also has bearing on genuine XRGs that have a central transverse swath of emission not traceable to either lobe. We point out that given the selection criteria our sample sources have predominantly long-winged I-dev sources as opposed to the more commonly seen "mini-winged" central inversion symmetric distortions in a large fraction of flux density limited samples \citep{Le1984}. The discussion below concerns only long-winged I-dev sources.

 Here we point out that while environment is known to play a strong role in shaping radio galaxy morphologies, most impressively exhibited by wide-angle tail and narrow-angle tail or head-tail radio galaxies (where ram pressure from the ambient ICM bends the lobes), and ram pressure could also affect lobes in O-dev and I-dev type sources that are located in rich galaxy groups and clusters, the rotationally symmetric distortions in O-dev and I-dev sources (in which the distortions are also centrally located) imply that ram pressure may not be relevant to their overall inversion symmetric characteristic.

Using the large sample we have imaged we are in a position to look at morphological clues to questions on suitability of some of the proposed scenarios for sources recognized as XRGs in the literature. For example, whether backflows have a role in forming wings and whether we can offer support for spin-flip scenario over and above what are only suggestions resulting from simulations.

\subsection{Proposed Model}
\label{s:proposed}

Considering only the objections to the main models suggested for XRG structures, namely the rapid spin-flip model and backflow deflection model, we note that on one hand the oft-stated inability of the deflection model to account for wings that are much larger than the main lobes and on the other hand the spin-flip model that runs into difficulties in explaining (1) the relic nature of wings Ð when relic lobes are recognized to be rare (2) the continuous and strongly bounded nature of the wings if they are just relics of past activity (3) inner edges and distinct emission gaps in the central regions in some XRG sources (referred to as I-dev sources here), and (4) why the wings are often directed along the minor axis of the host galaxy. 

If we accept that I-dev type sources with prominent wings as well as the ``genuine'' XRGs also result from an axis change, a rapid spin-flip of an active black hole over a large angle that then, after the direction flip, restarts as a new radio galaxy (of FR-II or relatively powerful FR-I type; option ``C'' in Figure~7 of \cite{DT2002}, \cite{M2002}) then the following scenario could overcome most of the objections of both models: backflows from the new lobes would, upon nearing the host galaxy, find oppositely directed channels of low-emissivity synchrotron plasma and will flow into them, thereby rejuvenating the old channels to then appear as well bounded, continuous wings that only reflect the flowing plasma. The rejuvenated channels, already previously excavated of the ISM or thermal halo or IGM material, can now have extents that can be larger than the main lobes: the wings will be enhanced if the backflows originate in a pressured region \citep{C2002} during the initial stages of the new radio source. 

{\em The remarkable finding that we have noted, that I-dev type offset emission occurs almost exclusively in FR-II type sources, is a strong support for models that require backflows to aid in the formation of the offset central wings.} Indeed, for most known XRGs as well, barring a few examples of XRGs with main radio sources of FR-I type \citep{SS2009}, such a property -- of main lobes having edge-brightened FR-II type morphology -- is recognized (e.g., \cite{C2002}). For the well-known XRG, NGC~326, \cite{HKR2012} have suggested, on the basis of analysis of radio, optical and X-ray datasets, the edge-darkened main lobes to have been of FR-II type which are now being observed at a stage where the beam activity has ceased and lobes have taken on an appearance of edge-darkened nature just as \cite{SS2009} made a case for B2014+56 and four other such edge-darkened XRGs. 

With the backflowing plasma flowing into the excavated channels, the wings do not any longer hold only the older relic plasma. The wings are therefore not any longer relic sources -- they will therefore not have the same limitations of visibility as relic radio galaxies. They will also not always exhibit the expected steep spectra of relic sources \citep{LRP}. The continuous run of emission between the main lobes and the off-axis emission or wings naturally results from backflows flowing into the past channels. 

The matter of emission gaps and sharp edges that have been cited as a problem for the spin-flip model is also solved if we understand them as resulting from buoyancy experienced by the old lobe plasma near the host galaxy location as also argued by \cite{HKR2012}. The central relic-lobe regions would over time then relocate away from the host galaxy due to buoyancy much like the X-ray holes in central cluster radio sources. The rejuvenated channels or wings can now have emission gaps as well as edges at their inner ends. 

\subsection{Unperturbed Black Holes and Host Minor Axes}

The one last obstacle to the spin-flip model -- the prevalence of minor axis wings in XRGs (and in sources with mini-wings in the 3CRR sample, see \cite{SS2009}) -- is a more formidable problem for which to account. Although we are yet to establish this relationship -- that of the wings extending along an axis close to the host minor axis for our entire sample of I-dev and genuine XRGs this has already been shown to be the case for 22 sources \citep{GCR16} and for samples of XRGs in the literature (many of which are I-dev sources; \cite{C2002,SS2009,HC2010,Paper0}). We suggest that the minor axis is a favored ``resting'' axis for black holes in radio galaxies, or for massive elliptical galaxies in general. If there have been no perturbations over a long period of time the black hole spin axis would continue to lie close to the host minor axis \citep{S12}. It is when there is a perturbation (such as a minor or major merger) that the black hole axis can change direction. If conditions are such that the flip occurs over a large angle and the AGN restarts its activity in the form of a powerful radio source, the resulting radio source morphology could be that of an XRG or an I-dev source with relatively large wings. So it is rather special circumstances indeed that could likely give rise to XRGs or I-dev sources with their impressive wing extents. 

Support for the ``resting'' axis idea lies in various radio-optical axis correlations noted between radio galaxies and their host elliptical galaxies: (1) \cite{C2002} for XRGs, (2) \cite{SS2009} for XRGs and Giant Radio Galaxies, FR-IIs, and 3CRR sources with central distortions, (3) \cite{BB2009} for FR-I sources, and (4) \cite{GCR16} for a sub-sample of our 87-source sample. While 3CRR FR-IIs are known to have their radio axes over a wide range of angles with respect to the host major axis both FR-Is (mostly lower power FR-Is) and giant radio galaxies (GRGs) are found to have radio axes predominantly close to the host {\em minor axis} \citep{BB2009, SS2009}. In addition, dust axis-optical axis correlations and dust axis-radio axis correlations for FR-Is, FR-IIs and GRGs have been understood within a framework where mergers needed to produce FR-IIs invariably contribute to disturbing the black hole axis where as FR-Is and GRGs are ``aided'' in their respective formation in more benign conditions not necessarily requiring mergers: Combination of lower power jets and larger host elliptical galaxies in the case of FR-Is removes the requirement for external fuel sources like mergers and as for GRGs, lower resistance to jet paths in the form of minor axis propagation and void regions on larger scales both provide conditions for unimpeded growth of jets to giant extents \citep{S12,MJSS}.

\subsection{A Combined Model to Understand I-dev Sources}

In the case of XRGs, the striking structures might be generated when the black hole experiences a perturbation resulting in a spin-flip (from the minor-axis direction) over a large angle (with axis now well away from the minor axis) and the AGN once again restarts its activity to then produce a radio galaxy with back flows capable of rejuvenating the old lobes. Just as GRGs may require a chance occurrence of jets that have formed along directions of least resistance -- host minor axes and regions devoid of neighboring galaxies \citep{MJSS} -- similarly XRGs and I-dev sources (with large wings) might be requiring a unique set of conditions to form. What the above scenario proposes is that {\em in creating XRGs and inner-deviation type sources (with large wings) both spin flip and  backflow deflection/escape (into old synchrotron plasma channels) are needed.} Such a ``combined'' model that we find persuasive and we suggest as a framework to explain our results for inner-deviation type sources is also what is suggested by \cite{HKR2012} for accounting for the structure of NGC~326. 

With the variety of imaged source structures in I-dev sources how do we understand them in the light of the above model? We used the illustrative chart in Figure~\ref{fig:chart} to see how structures may be affected by two of the parameters -- time since spin-flip when the new activity started and age of the new activity, for cases where the spin-flip happens in the sky plane. On the lower left corner of the chart one expects well-formed and well-bounded "wing" sources to be associated with physically small-size main sources. As these new sources age and grow in size (and hence shift to the lower-right) the wings will be affected by synchrotron as well as expansion losses, therefore appearing less bounded. To check if main-lobe linear sizes in the lower half of the chart are on average in the expected sense rather than select sources on the basis of "well-formed and well-bounded" wing structures, we selected on one hand, all those sources with {\em both} wings having angular extents as large as or larger than the main lobes and another group with wings with angular extents smaller than the main lobes (with firm redshifts). We find that for the four sources in the former group (only one source is a quasar) the average linear size of the main source (142~kpc) is less than half that of the group of twelve in the latter group (389~kpc). 

This trend of larger radio galaxies tending not to have proportionately large wings was already shown in Figure~5 of \cite{SS2009}. We can understand this division -- of the ratio of wing to main lobe extent -- with linear size of the main source within the combined model using the plasma over-pressure model of \cite{C2002}: in the initial stages after the spin-flip as the young jet makes its way along the major axis direction it faces impedance and lobe plasma pressure will build up, which at some stage will escape and expand in a direction closer to the minor axis direction and into the channels of the relic lobes -- resulting in a XRG. As the active jet continues to grow to sizes well outside the host galaxy however there will be less such pressured plasma available for the continued rejuvenation of the old lobes. On the other hand backflows that accompany powerful lobes will help in rejuvenating the old lobes although their efficacy is being questioned as a result of turbulence that can set in \citep{HKA}. Our preliminary finding of stronger hotspots located in lobes associated with longer wings supports the role of backflows in creating the central wings. As sources age the wings will also become less visible as they suffer synchrotron and expansion losses.

The question, therefore, of the inability of sub-sonic synchrotron flows in giving rise to wings as large or larger than the main radio galaxy arises more urgently only for smaller-size radio galaxies, as indicated above. In such sources the existing wings (already likely to be larger than the main radio source) that accompany the as yet small-sized new source will be further assisted by the escape of pressured plasma. 

\subsection{Inner Emission Gaps}

The emission gaps and edges between the inner ends of lobes would start becoming visible in sources in all remaining quarters of the chart. Such wings are more likely to be less bounded and relatively fainter (than the wings of small-size radio galaxies populating the lower-left corner). The top left corner will be populated by physically small size sources with rejuvenated (old) wings and ones that are more likely also accompanied by the Z-shape. As such sources grow in size (and thereby occupying the top-right corner) they will be accompanied by distinctly Z-shaped but faint and frayed wings. Of course simulations of the combined spin-flip, plasma over-pressure and backflow model would need to be carried out for combinations of quiescent time and age of the new episode to understand the types of low axial ratio source structures such as what we see in our sample. Our present observations form a useful resource to explore, via simulations, consequences of axis rotations to source morphologies and hence infer axis rotation types and parameters. 

If I-dev type sources  with long wings  also suffer axis changes then along with O-dev sources a non-negligible fraction of sources change from being QSOs to non-QSOs and vice versa (for those where the flip or precession is by a larger angle than 45 deg).

The suggestion that I-dev type source structures  with long wings may be resulting from axis rotation even as the jets are active as opposed to the above case where a spin flip occurs between two separate activity episodes (option ``B'' in Figure~7 of \cite{DT2002}; see also \cite{Go2011}) implies that in those identified with QSO hosts one should see effects of projection relevant to axis rotation such as non-collinearity, just as in the quasar O-dev sources discussed above. There are six I-dev sources identified with quasars three of which display non-collinearity. The non-collinearities are not strong nor are signatures of projection. Only for one source, J1206+3812 the weaker and closer hotspot is displaced to the same side as its associated lobe off-axis emission.  Extending this to the larger sample of I-dev sources will, of course, be useful but with the few quasar I-dev sources examined it appears that axis precession of an active jet while not ruled out may not be the primary cause of  such I-dev morphologies.

\section{SUMMARY}
\label{s:Summary}

Our analysis of the total intensity images of the entire sample of 100 XRG candidates that we describe as ``low-axial-ratio'' sample has allowed us to map the off-axis emission more clearly. In this paper, the second in the series of papers dealing with our VLA observations, we examine the images to know the phenomenology of how the off-axis emission relates to the main radio source. With detailed structures now available for most of the 100 sources we are beginning to unravel patterns to the relationship of secondary off-axis emission with their primary radio doubles, which also allows us to explore and present frameworks for the origin of this emission that is located significantly offset from the main radio axis. 

The sample and multi-band continuum data we have gathered provide a useful resource to explore black hole axis changes. With the rich haul of sources that have likely undergone axis rotation or spin flips our work has shown the usefulness of the approach of looking at low axial ratio radio galaxies to pursue searches for binary supermassive black holes.  The rotationally symmetric morphologies seen in O-devs suggest an axis-related cause and hence a phenomenon related to the central black hole. An axis related cause is also suggested in the case of XRGs and in I-devs with large wings (that are XRGs with emission gaps) in one set of major models \citep{M2002}. Perturbation to the central black  axis has been studied in relation to binary black holes that result from galaxy mergers \citep{Be1980,M2002}, although ``realignment'' phenomena \citep{NP} as well as disk-related phenomena too have been implicated \citep{KP}. With galaxy mergers expected and known to happen and with growing examples of binary AGN in galaxies (e..g., \cite{MS}), binary black holes are a strong contender in causing axis changes. 

In the present work, we also invoke role of backflows in XRGs and I-devs with large wing extents in assisting growth of the transverse wings, using a so-called combined model, although it is to be noted that backflows are always expected in powerful radio galaxies (and are expected to create central inversion symmetric wings under favorable conditions of asymmetric host halo encounters).

In a sample chosen on the basis of  the presence of  substantial off-axis emission our observations have revealed the following:
\begin{enumerate}
\item Off-axis emission structures are invariably inversion symmetric. 

\item The inversion symmetric off-axis emission in radio galaxies appears to have origins from strategic locations, either outer ends of radio lobes or their inner ends, confirming results we presented in \cite{Paper0}. 

\item The latter type of off-axis emission features originating from inner ends of lobes occur only in FR-II type radio galaxies whereas the former type -- originating from the outer ends of lobes -- can form in both FR types. 

\item Off-axis emission originating from inner ends of lobes are the most common, dominating off-axis emission originating from outer ends of lobes by two to one. 

\item A sample of twelve sources with clear S and Z-shape are identified. If O-dev sources (that also include all S and Z-shaped sources) represent source morphologies resulting from black hole axis changes then a minimum of one percent of radio galaxies undergo axis rotations. 

\item We recognize expected signs for rotating axes in O-dev quasars. This shows O-dev structures in general to be a strong signature of rotating axes in radio galaxies.

\item Physically large radio galaxies do not form proportionately large wings unlike physically small size radio sources.

\item X-shaped radio galaxies are suggested as being subset of the I-dev type sources. 

\item We offer a scenario for the creation of  I-dev type sources with impressive wing extents. Impressive length orthogonal wing radio galaxies appear to form under rare circumstances, with one of the more persuasive scenarios being a combined model of two of the leading and contending models for XRGs in the literature, where backflows from a later, powerful activity flow into relic channels of past activity that were oriented in a nearly orthogonal direction. Not only is a large angle axis flip followed by a new and powerful activity needed but the orientation of the host galaxy and the new jets is such that backflows can deflect past the host in inversion symmetric manner.  Backflows appear to be crucial in creating the smooth wing structures as well as in keeping the relic channels bright and in creating impressive wing extents that can be longer than the main lobes. Time can affect the appearance of the winged radio source and introduce emission gaps; Idevs, with their separated wings may be older. 

\item Including I-dev sources, at least 4\% of radio galaxies appear to undergo black hole axis rotation. 

\item With features such as non-collinearities, central inner-S ``spine,'' corresponding inner emission peaks in the two lobes, double hotspots as well as protruding hotspots common in the sample sources low axial ratio samples form a good resource to search for binary blackhole candidates \citep{D2014}. 

\end{enumerate}

Based on the structures we have imaged and categorized we have offered insights into likely processes that could be generating the seen structures. The data we have 
gathered will be used in testing these frameworks. Obviously the parameter space we are exploring is large. We urge simulations of axis change phenomena to fully make 
use of the gathered data to reproduce the structures and identify the different axis change types occurring in radio galaxies.

\section{FUTURE INVESTIGATIONS}
\label{s:Future}

The total intensity multi-band continuum imaging of our sample was carried out for the specific purpose of revealing clues regarding nature of the off axis emission that forms in a fraction of extragalactic radio sources. The present work has shown a pattern to these off-axis emission regions -- the strategic manner in which they link with the main radio source and together with the FR type of the main source how they provide hints regarding the physical processes that could generate them. Our future studies will include a full polarization and spectral index characterization of our sample sources as well as more detailed analysis of individually intriguing sources. We will present and discuss the polarization and spectral index information on our source sample and separately will use all the information we have gathered on the sources to derive timescales for axis perturbations and also the viability of suggested models for off-axis structures. The data we have gathered and analyzed will be crucially served by including optical imaging and spectroscopic observations of our sample sources. 

 Searches for binary black holes are a growing research area today. Techniques used in the searches combine two or more AGN indicators such as double peaked optical spectral lines,  twin radio-core detection, X-ray source detection, and radio galaxy morphologies such as S or Z-type morphologies (e.g., \cite{MS}). Our sample sources with their  S-, Z- and X-shaped structures are ideal for searches for binary AGN. We are pursuing high-resolution optical imaging and spectroscopy to look for presence of multiple optical nuclei  and obtain spectra for hosts that do not have spectra, in addition to VLBI observations of strong core radio sources to search for presence of multiple radio cores.


\section{ACKNOWLEDGMENTS}

The National Radio Astronomy Observatory is a facility of the National Science Foundation, operated under cooperative agreement by Associated Universities, Inc.  Funding for the Sloan Digital Sky Survey IV has been provided by the Alfred P. Sloan Foundation, the U.S. Department of Energy Office of Science, and the Participating Institutions. SDSS-IV acknowledges support and resources from the Center for High-Performance Computing at the University of Utah. The SDSS web site is www.sdss.org. We thank Ravi Subrahmanyan for very useful conversations. D.\ H.\ R.\ gratefully acknowledges the support of the William R. Kenan, Jr.\ Charitable Trust.

\section{APPENDIX -- NOTES ON INDIVIDUAL SOURCES}

With all available maps at hand we have tried to present a plausible scenario for at least some of the sample sources. The inversion symmetry, the central transverse extensions, the different source shapes whether of S or Z type, the link seen of the off-axis emission with strategic locations of the source as well as connection with the host (whether radio-optical relation or geometrical relation or host type) all give clues to how the source structures may have formed. The following notes highlight morphological aspects based on our VLA images and present reasoning for inferences drawn. For more information on these sources particularly related to optical data the reader is referred to the comprehensive notes in \cite{C2007} and also \cite{C2009} and \cite{L10}.

\begin{enumerate}

\item J0001$-$0033: Two radio sources at a small angle to each other. Resembles 3C\,293 and J0707$-$326 \citep{Sa2013} although there are differences. In J0001$-$0033 as well as 3C\,293 in addition to two separate sources one sees O-dev type off axis emission from the lobe ends. The O-dev type off axis emission suggests a trail left behind by the moving axis, which rotates even as the jets may be actively transporting synchrotron plasma. So we may interpret the image as one where the axis may have spent sufficient time to create each of the sources and it may have experienced a {\em perturbation during the activity.} Where as in J0707$-$326 the axis may have experienced a {\em perturbation when it was inactive after having created one source to then create a new source in the changed direction.} The extended source in J0001$-$0033 shows curved or helical features in the NE lobe. Could the jet have experienced a small-angle precession before being disturbed to the EW direction? An Inner-S structure is seen in the source.

\item J0036+0048: This is a quasar with prominent lobes and prominent off-axis emission with origin near the inner ends of the two lobes.

\item J0045+0021: Prominent wings with origin near the inner ends of lobes.

\item J0113+0106: This is a giant edge-brightened radio galaxy with prominent central wings originating from the inner ends of the lobes. The wings are nearly orthogonal to the source axis and are wide, contrasting with the rather narrow collimated lobes. The hotspots in each lobe are accompanied by rather compact, single emission peaks. The two companion emission peaks lie on opposing sides of the radio axis formed by the main hotspots and core. 

\item J0115$-$0000: Short central wings linked to the inner ends of lobes. The two wings do not share the same position angle.

\item J0144$-$0830: With no radio core, jets or hotspots this source is likely to be a relic radio source. The BS map reveals a transverse extension along the central region of the source on both sides of the source although more extended to the south. We classify this source as a XRG candidate where the black hole axis may have undergone a rapid large-angle flip between two activity epochs.

\item J0145$-$0159: The off-axis emission in this FR-I source originates from the outer ends of the source in opposite directions. This off-axis emission is seen to extend all the way back along the axis and towards the central regions reminiscent of backflows in FR-II sources. The emission stops around the central regions and even shows transverse emission particularly in the CS map. An explanation for this morphology might be a slow axis change that initially commenced in almost EW direction and then stopped in the present NW direction while all the time remaining active. The available polarization map clearly shows a 90-degree shift in going from the off-axis emission to the transverse extensions where the projected B field is seen aligned along these extensions. The similarly bright galaxy to NNE of the host (that also hosts a compact radio source) has the same redshift and lies 26 Kpc away from the host.

\item J0147$-$0851: This edge-brightened source has only one prominent wing that is linked to the inner end of the SW lobe. However in the FIRST image faint diffuse off-axis emission associated with the NE is seen. The two hotspots are not collinear with the radio core. A peculiar feature is seen adjacent to the SW hotspot that is elongated along the same axis as formed by the NE lobe. The nature of this feature and its connection to the edge-brightened radio source is unclear.

\item J0211$-$0920: Although edge-brightened the two lobes appear relaxed and do not have compact hotspots at the lobe ends. The two off-axis emission regions that share similar position angle are themselves broad and their links to the main lobes are not clear, whether to the inner or outer lobe ends. 

\item J0225$-$0738: an FR-II source with I-dev type off axis emission although seen clearly only for the NW lobe (a feature that is longer that its lobe extent and nearly orthogonal to it). The SE lobe has a bright low-axial ratio lobe with no distinct off-axis emission feature in the opposite direction that originates from the inner-lobe region but instead has emission that is asymmetrically distributed to the west of the radio axis formed by the hotspot and the core. A feature of the source is lack of collinearity and presence of double hotspot in the NW lobe and likely double hotspots in the SE lobe (indicated by the broad lobe head).

\item J0702+5002: The striking feature of this I-dev source is the pair of prominent transverse wings that have extents much larger than their respective lobes. The lobes of the source are moreover of FR-I type. 

\item J0725+5835: Similar to 0225$-$0738 where one lobe (SE lobe) has a distinct off-axis feature that is nearly orthogonal to the lobe axis. This feature has a sharp bend to the SE at its end. The compact feature in this wing is most likely an unrelated radio source. It has a similarly faint optical ID as the radio core. The NW lobe although along a position angle same as the SE lobe. it is displaced to the east and is associated with the short off-axis, central emission feature extending in the opposite (SW) direction. 

\item J0805+4854: Only one transverse, central wing is seen in this source that is associated with the SE lobe. This wing is broad and of relatively short extent. A mild inner S-shape is recognized in this bright core FR-II.

\item J0813+4347: The structure of the source is complex and difficult to classify. An inner compact triple source with a bright core, straddled by two peaks of emission, extends to the east into a broad edge-brightened lobe with its emission more prominent to the south. There is no obvious hotspot associated with this broad outer lobe. To the west the inner emission peak (of the inner triple source) is directly linked to a broad feature that extends to the north. There is a likely inner S-shape that is surrounded by emission with an overall Z-shape. The inner triple may also be indicative of a likely restarted activity.

\item J0836+3125: The northern lobe with its double hotspot is not collinear with the axis formed by the core and hotspot in the opposite lobe. Instead, a likely feature to the SW of the double hotspot forms an axis with the core and the hotspot at the end of the opposite lobe. Both the lobes are prominently to the east of this axis.  A wing is associated with the northern lobe originating at its inner end close to the core. The image does not reveal any wing associated with the southern lobe.

\item J0838+3253: A clear Z-shaped (hence O-dev type) source where a core and two straight jets connect with edge-brightened lobe regions that have weak trailing emission in opposite directions which extend along distinctly different position angles. The northern trail includes an unrelated compact radio source that coincides with a galaxy. The two jets are not smooth and show multiple (possibly corresponding) knots.

\item J0845+4031: A clear S-shaped (hence O-dev type) edge-brightened source. The core is straddled by a curving string of extended knots that connect to edge-brightened lobes each with a hotspot at its leading end. The lobes themselves leave trails of weaker emission in opposite directions. 

\item J0846+3956: This is a typical I-dev source where the off-axis emission features originate from the inner ends of lobes. In addition to the two central wing features both lobes also have inner elongated emission features that are offset by a small angle from the main radio axis in opposite directions. The source has an inner-Z shape similar to J0707-326 \citep{Sa2013} although in the latter it is more pronounced.

\item J0859$-$0433: In both lobes the (pronounced) off-axis emission features start to deviate sharply from the main axis well before the core region. The deviation points are represented with two emission peaks, which are at similar distances from the core. The hotspots in the two lobes are bright and are located at quite different distances from the core in the asymmetric radio galaxy. There is an optical object almost exactly around the inner peak in the east lobe where the lobe veers sharply to the SW. What may be the inner pair of emission peaks signifying? Could the axis have drifted from nearly N-S direction all the way to the inner emission peaks where the backflows from the inner peaks got deflected into the older trails? After creating the inner source the axis may have marginally rotated towards E-W to then create the main (protruding) hotspots. The SW off-axis emission extends well past the core unlike the opposing wing. 

\item J0914+1715: This FR-II source has very prominent and extended wings that originate from the inner ends of lobes. Besides the two strong hotspots at the leading ends of the two lobes there are inner emission peaks, one in each lobe, that separately form an axis with the radio core displaced by a small angle from the main axis.
A possible scenario for this source might be that the axis may have drifted or precessed from a NE-SW direction to form the inner source represented by the two inner emission peaks after which it may have precessed again by a small angle to create the two strong and protruding hotspots.

\item J0917+0523: The E lobe morphology is similar to the E lobe of 0859-0433. The associated off-axis emission veers off sharply away from the axis well before reaching the core vicinity. Because of a bright star close to it in this case we may not know if there is an object in its path causing this sharp bend. Separately though it is interesting that the inner peak location where the sharp bend occurs is collinear with the core and W hotspot.

\item J0924+4233: The source, with its edge-brightened and relatively relaxed, low-axial ratio lobes is accompanied by broad inversion-symmetric off-axis emission features that appear to originate from different lobe locations on the two sides. For the SW lobe the off-axis feature appears to originate from region closer to the inner section of the lobe whereas for the NE lobe, from the outer end of the lobe. The L-band image shows a likely restarting source where the host is straddled by a close pair of emission features.

\item J0941$-$0143: This edge-brightened source has a prominent wing associated with its NE lobe. It originates from the inner end of the lobe and has an extent more than the extent of its respective lobe. However for the SW lobe there does not appear to be a corresponding wing although a feature is seen extending beyond the leading end of the SW lobe and displaced to the side opposite to the wing linked to the NE lobe. A possibility is that the wing associated with the SW lobe is seen in projection.

\item J0941+2147: This edge-brightened source has one off-axis wing to the NW,  associated with the southern lobe. 

\item J0943+2834: This is a typical edge-brightened source with off-axis emission originating from inner ends of the two lobes. 

\item J1005+1154: The classification of this highly asymmetric edge-brightened source is unclear. The two lobes have relatively low axial ratios and the off-axis emission features are also broad making it difficult to locate the origin of these features. For the northern lobe the off-axis emission may be likely associated with the inner lobe end.

\item J1008+0030: this source may have undergone an axis drift. The ``halo'' may just be the old emission in a different axis. Inner-S shape is clearly seen.

\item J1015+5944: This is an edge-brightened source with one prominent wing, associated with the NW lobe. A short off-axis emission is seen associated with the SE lobe in the S-band, B-array image.

\item J1040+5056: Relaxed lobes. O-dev type transverse extensions seen right near lobe ends although these off-axis extensions also extend all the way down to regions closer to the central regions of the source. Shows an inner-Z and a slight inner-S shape. Likely a restarted source because of its bright and extended core.

\item J1043+3131: The source structure does not lend itself easily to classification into the three categories we define in this paper. With the data available with us however we can suggest a scenario for its formation. The host galaxy is in the vicinity of two other similarly bright galaxies one on either side of the host along a near N-S direction; optical spectra are not available for these neighboring galaxies. The host and its northern neighbor have similar r-magnitudes and assuming the same redshift as for the host it is about 3~kpc away. Our VLA data have allowed us to derive spectral index distribution for this source. The radio halo is found to have a layered or nested spectral index distribution that also correlates with the total intensity distribution in the halo; this appears to reveal newer plasma within older plasma in the radio halo. A bright radio core is accompanied by a jet to the south with a compact hotspot at its leading end and forms a straight axis with a compact hotspot on the opposite side. A possible scenario for the source is one where the original activity was in the orthogonal (NE-SW) direction after which the AGN axis may have suffered a large angle, nearly 90-deg flip to then create the new source with compact hotspots. The small linear extent of the source suggests the older relic plasma may be confined in the thermal halo of the host and its close neighbor through which the new jets make their way. With a BLRG host spectrum the SE jet may be the approaching jet. The backflow from the jets emerging out of the halo (possibly the reason for the protruding southern hotspot) would add younger plasma to the halo therefore contributing to a layered spectral index distribution. Simulations will be needed to test out these ideas. The NW chimney-like feature continues to be a puzzle.

\item J1049+4422: The asymmetric edge-brightened radio galaxy has I-dev type off-axis emission associated with one of its lobes, the NW lobe. For the opposing SE lobe the inversion-symmetric off-axis feature is wide and associated with the entire lobe making it difficult to classify it as I-dev or O-dev type. It is likely that we see an I-dev type off-axis emission in projection like in the source, 0941-0143. 

\item J1054+5521: Lack of a radio core, absence of an optical object in between the two lobes as well as presence of unrelated point sources in the lobes makes it difficult to easily recognize the nature of this source. The B-array, S-band map shows two edge-brightened lobes (with east lobe having a bright and compact hotspot at its leading end). Inversion symmetric off-axis features are seen extending from the inner ends of the two lobes. 

\item J1055$-$0707: A typical edge-brightened radio galaxy with relatively short inversion symmetric features extending from the inner ends of the lobes. 

\item J1102+0250: Prominent inversion symmetric O-dev type off-axis emission features accompany this Z-shaped, edge-brightened radio galaxy that has a bright core and a prominent continuous (and knotty) jet to the NW and a likely jet to the SE. Both off-axis features have broad emission peaks that share an axis with the radio core displaced from the main radio axis by a small angle. 

\item J1114+2632: A bright core straddled by asymmetrically separated hotspots which form the origin points for wide, O-dev type off-axis emission in opposite directions. It has a S-shape to its extended morphology and the main axis also has an S-shape. Although optical spectrum is not available for the faint host the radio structure hints at a quasar-type nature.

\item J1120+4354: This edge-brightened source is classified as an XRG given the central swath of emission orthogonal to the source axis and not linked to either the inner or outer ends of lobes. The source has double-hotspot features at each lobe extremity.

\item J1135$-$0737: This is an edge-brightened Z-shaped source with wide off-axis emission features that span the length of the two lobes. 

\item J1140+1057: This edge-brightened source with a bright core has clear off-axis emission features extending in opposite directions from the inner ends of the two lobes. Both lobes exhibit double hotspots. A curious short, jet-like feature is seen almost orthogonal to the main source axis to the east of the core.

\item J1200+6105: Our higher resolution image while confirming the inference in \cite{C2007} does not make it any clearer as to the missing ``wing'' associated with the eastern lobe. The inner structure of the source shows hints of changing axis: in addition to the larger-scale inner S-shape along the source axis there is an elongated feature across the core at nearly 40-deg from the main axis. Both lobes contain inner emission peaks in addition to the bright hotspots at the leading ends of the source. There is only one wing associated with the source; it is not clear which lobe it is linked to.

\item J1202+4915: A typical edge-brightened source with off-axis emission originating from the inner ends of the lobes. Both lobes have extended emission peaks at the leading ends suggestive of double hotspots or twin emission peaks. 

\item J1206+3812: A typical edge-brightened source, a quasar, with off-axis emission originating from the inner ends of the lobes. 

\item J1207+3352: An edge-brightened source with a bright core that shows an S-shaped spine accompanied by broad off-axis features originating from the outer ends of the lobes. Both lobes show sharp, narrow features (more prominent in the northern lobe) extending nearly orthogonal to the source axis. 

\item J1210$-$0341: The edge-brightened source has off-axis emission linked to each lobe that also forms an orthogonal swath across the host galaxy. These off-axis emission features span the entire length of the two lobes; the lobe axes are not collinear. There is a hint of a jet connecting the core to the SE hotspot (the L-band archival image). Our new JVLA S-band B-array image reveals more extended emission related to this central swath, particularly to the south-east. From our polarization images we note that the B-field vectors while uniform over the main lobes dramatically change directions orthogonally at the ``leading'' ends of the central emission swath to indicate circumferential fields. The source structure could be indicating a case of relatively slow change in jet direction (over a large angle) in contrast to the ``rapid flip'' invoked in XRGs.

\item J1211+4539: A typical edge-brightened source with short off-axis emission originating from the inner ends of the lobes.

\item J1218+1955: A narrow edge-brightened radio galaxy with short off-axis emission features originating from the inner ends of the lobes. The two inversion-symmetric off-axis features form a central swath (although not orthogonal) across the host galaxy.

\item J1227$-$0742: This is again an edge-brightened source where the leading ends appear to protrude out of a pair of broad and inversion-symmetric lobes. The morphology is similar to that of 0707-326 \citep{Sa2013} and suggests an axis change that could have occurred between epochs that created the inner relic source and the more recent edge-brightened source where the new jets emerge out of the relic plasma creating a bottle-necked structure.

\item J1228+2642: This edge-brightened radio galaxy has compact hotspots at the leading ends; the SW hotspot is particularly point-like and also more protruding. Both lobes also exhibit inner emission peaks on either side of the source axis. The source lacks wings although we recognize mild inversion symmetry of the lobes. The main axis of the source is along the host minor axis. We classify this source as one which may have suffered a small-angle axis-change. 

\item J1232$-$0717: The poor resolution of the image prevents a clear classification of the source, however we recognize an asymmetric edge-brightened source with a central S-shape spine that has unusually broad lobes. The lobes include two emission peaks that form a separate axis with the central core. We classify this source as 
one which may have suffered a large-angle axis-change.

\item J1247+4646: An edge-brightened source with a mild inner S-shape accompanied by a central swath of emission orthogonal to the source axis. 

\item J1253+3435: This is a highly non-collinear edge-brightened source that has at the leading ends of its wide lobes, a double hotspot in one and an extended hotspot in the other. Both lobes extend off-axis in opposite directions over their entire length, lending it Z-type morphology. 

\item J1258+3227: This is a narrow edge-brightened radio galaxy with a single off-axis emission feature linked to the inner end of the NW lobe. Although a compact hotspot is seen at the leading end of only one of its lobes, the NW lobe, the structure of the SE lobe does not rule out a hotspot at its leading end. We classify this source as an I-dev type source based on the clear central wing associated with its NW lobe. 

\item J1309$-$0012: This edge-brightened radio galaxy has a prominent wing associated with its western lobe at its inner end. There is no such wing seen associated with the eastern lobe. However the FIRST image does hint at a weak, central off-axis emission associated with the western lobe.

\item J1310+5458: Our B-array, S-band image shows a clear S-shaped source with broad lobes and an inner-S-shaped spine connecting the two hotspots in this edge-brightened quasar. 

\item J1316+2427: This edge-brightened source is an I-dev type source with short off-axis extensions that originate from the inner ends of lobes.

\item J1327$-$0203: This is an edge-darkened source with an inner pair of broad emission peaks straddling the core that has emission extending beyond them along the radio axis. Although this source exhibits an overall X-shaped morphology with its central swath of emission not obviously traced to either of the lobes we recognize a central Z-type structure. A scenario we offer for this source is that it began as an O-dev type source where the axis would have drifted over a large angle to create the inner source, which since has ceased activity. The source then may have restarted its activity along the same axis to produce the protruding emission extensions beyond the inner lobes.

\item J1342+2547: This is a non-collinear quasar, which has broad, inversion-symmetric off-axis emission features that span the lobe extents. 

\item J1345+5233: This edge-brightened source has protruding hotspots at the leading ends of its lobe with each lobe also possibly linked to centrally located off-axis emission features. Higher resolution EVLA data is needed to unravel more about its structure.

\item J1348+4411: This is a highly non-collinear edge-brightened source. However the non-collinearity almost disappears if we consider the inner peak in the northern lobe (seen more clearly in our archival VLA L-band image). The source appears to have a S-shaped spine linking the outermost hotspots with the northern inner peak and the core. Both lobes are linked to prominent off-axis emission features although for the northern lobe it originates from the outer end of the lobe where as it is not so clear for the southern lobe where the off-axis feature spans the entire length of the lobe.

\item J1351+5559: This is a clear FR-I type S-shaped source that shows signs of projection although the host galaxy does not reveal any broad emission lines or even strong narrow lines. A neighboring galaxy to the north is associated with a compact radio source that is superposed on the northern part of the S-shaped source.

\item J1353+0724: Please refer to notes in \cite{C2007}.

\item J1406$-$0154: This asymmetric, edge-brightened source has protruding hotspots at the leading ends of the lobes. Each lobe is associated with an off-axis emission feature originating from the inner lobe end for the E lobe whereas it appears to span the entire lobe length for the western lobe. 

\item J1406+0657: This is an edge-brightened XRG where the central swath of emission is along an orthogonal direction to the source axis. Please refer to notes in Roberts et al (2015a) and Cheung (2007) for more information.

\item J1408+0225: The halo-type structure of this source, identified with a blazar does not allow easy classification. However we recognize the presence of several emission peaks in this source besides the central radio core and a possible jet-like feature to the west. We also recognize a central S-type structure to the source. Please refer to notes in Roberts et al (2015a) and Cheung (2007) for more information.

\item J1411+0907: The suggested optical host has a close companion to the SE having a slightly fainter magnitude. Without a radio core detection we are unable to confirm which of the two is the likely host of the radio source. 

\item J1424+2637: This edge-brightened source has recessed extended emission peaks along an inner, mild-S-shape spine. The lobes have inversion-symmetric off-axis extensions spanning a large length of the lobes. We classify this source as O-dev type. 

\item J1430+5217: This quasar exhibits several interesting features. Not only do the two inversion-symmetric off-axis emission features clearly originate from the outer extremities of the source lending it a Z-shape but the inner spine formed by separated knots has a mild-S-shape. Apart from the one-sided knot like jet feature to the NW the source shows several pairs of emission peaks on the two sides each of which shares a separate axis with the central core.

\item J1433+0037: Please refer to notes in \cite{C2007} for more information. This source shows similarities with 0941-0143 and 1049+4422 where one lobe has an off-axis emission feature whereas for the opposing lobe no such feature is seen although weak extended emission is seen beyond the leading end. This is suggestive of likely projection effects acting on the putative off-axis feature associated with that lobe.

\item J1434+5906: This edge-brightened source appears to have a broad one-sided off-axis emission feature to the SE matched only by a broad protrusion on the opposing side. The source exhibits an inner S-shape spine connecting the two protruding hotspots and the core. 

\item J1437+0834: Please refer to notes in \cite{C2007} for more information. This edge-brightened I-dev type source with its short wings exhibits an inner-S-shape in the only available (FIRST) map.

\item J1444+4147: In the low resolution C-array, S-band image this edge-brightened source is seen associated with prominent central swath of off-axis emission that however can be traced to the inner ends of individual lobes. Higher resolution data would help to trace the wings better.

\item J1454+2732: This source requires more data to understand its morphology beyond the edge-brightened lobes with off-axis emission seen in the FIRST image. Given the location of the prospective optical host we classify this source as a likely O-dev type off-axis structure. 

\item J1455+3237: This is a relaxed but edge-brightened source with a mild inner S-shape. The low axial ratio lobes show broad extensions in opposite directions although this is more obvious for the SE lobe.

\item J1456+2542: This edge-brightened source appears to be a relic source with inversion-symmetric off-axis extensions associated with each lobe. However better sensitivity and resolution is needed to trace how they relate to the two lobes.

\item J1459+2903: This edge-brightened source has relaxed lobes with broad inversion-symmetric extensions associated with the inner ends of the individual lobes. Our L-band, A-array  archival VLA data reveals short twin extensions on either side of the core, suggestive of a restarted activity.

\item J1501+0752: In this edge-brightened source while the southern lobe has a short wing associated with its inner end the northern lobe exhibits a more extended feature that originates from the mid-section of the lobe in the opposite direction. As in J1210-0341 the two edge-brightened lobes are not collinear, offset towards the same side as their respective off-axis emission features.

\item J1515+0532: This narrow edge-brightened radio galaxy shows inversion-symmetric short extensions originating from the inner ends of its lobes.

\item J1537+2648: This edge-brightened source has dissimilar lobes. The NW lobe is a typical FR-II type lobe with a hotspot at its leading end and is associated with an O-dev type off-axis emission feature. The SE lobe however is almost rectangular and shows a narrow linear feature running along its western edge and terminating in an emission peak at the end. No off-axis emission is seen associated with this SE lobe. Projection effects may be affecting the lobe structure.

\item J1600+2058: The edge-brightened source is a clear I-dev type source with inversion symmetric off-axis emission features originating from the inner ends of the two lobes. The lobes although of edge-brightened nature exhibit a central spine of emission that shows prominent curvature in the northern lobe and connecting the recessed hotspot and the core. The core itself is extended and connected to a jet-like feature to the NE.

\item J1603+5242: This is an edge-brightened quasar with non-collinear bright hotspots at the leading ends and with lobes that exhibit I-dev type short off-axis emission features. The slightly extended core is suggestive of an associated jet to the SE. The protruding nature of the two hotspots and a likely jet oriented away from the SE hotspot may be indicative of changes in the jet axis of this quasar.

\item J1606+0000: This unique source has been studied in great detail by \cite{HKR2010}. Besides revealing all the features seen in the VLA images presented by \cite{HKR2010} our JVLA data also shows the curved nature of the northern plume linking it to the extended emission that we map over a wider region than previously mapped. The central jet itself shows a Z-shape over the extent of the inner source. The extended plumes are mapped more robustly and suggest the possibility of a much more extended (edge-brightened?) radio source along the axis formed by the two plumes and the radio core. The emission extending in opposite directions about the plumes suggests that the axis may have rotated. The jet axis may have rotated over a large angle to create the more recent small inner radio source with the jet towards the NE. 

\item J1606+4517: A likely relic but edge-brightened source with I-dev type off-axis emission features.

\item J1614+2817: An edge-brightened source with wide lobes and a wide central swath of emission that is nearly orthogonal to the radio axis.

\item J1625+2705: This quasar presents an interesting structure where neither the two hotspots are collinear with the core nor is the core and its SE one-sided jet collinear with either of the hotspots. The jet exhibits a sharp bend at its end similar to but inversion-symmetric to the bend seen at the hotspot in the opposite lobe. On larger scales there is short but wide emission seen to the east orthogonal to the source axis whereas to the west of the quasar is seen weak diffuse emission about 20 arcsec away. With the data we have it is difficult to interpret and classify the source structure although the non-colinearities, the inner-S, the inner Z-shape and the faint orthogonal extensions suggest the quasar axis may have undergone changes in direction.

\item J1653+3115: This is a clear FR-I type and S-shaped quasar.

\item J1655+4551: This edge-brightened source exhibits only one wing to the west. There are indications that it is associated with the NW lobe originating from its inner end. The SE lobe has a bright hotspot like feature well recessed from the lobe end located closer than half way to the lobe end. Could the wavy extension well beyond the recessed hotspot be a fortuitous projection on to the lobe of a prominent wing associated with the inner end of that SE lobe, much like suspected in sources such as J0941-0143, J1049+4422 and J1433+0037?

\item J1656+3952: This is a clear Z-shaped source associated with an edge-brightened radio galaxy.

\item J2359$-$1041: In this edge-brightened source there are several emission peaks. There is a curved, jet-like feature through the core connecting it to hotspots at its extremities (seen more clearly in the B-array L-band map). Besides this there are two outer emission peaks, which lie on opposite sides of the inner S-shaped spine. 

\end{enumerate}


\end{document}